\documentclass[oldversion]{aa}  
\usepackage{graphicx,psfig,amsmath}

\newcommand{\bp}{$\beta$\,Pictoris}
\usepackage{color}

\begin{document}

   \title{Fe\,I in the \bp\ circumstellar gas disk}
   \subtitle{I. Physical properties of the neutral iron gas}
   \author{
   A.~Vidal-Madjar\inst{1,2}      
   \and
   F.~Kiefer\inst{1,2,3}
   \and  
   A.~Lecavelier des Etangs\inst{1,2}
   \and 
   V.~Bourrier\inst{4}
   \and
   D.~Ehrenreich\inst{4}
   \and
   R.~Ferlet\inst{1,2}
   \and
   G.~H\'ebrard\inst{1,2}
   \and
   P.~A.~Wilson\inst{1,2}
 }
   
\authorrunning{Vidal-Madjar et al.}
\titlerunning{Fe\,I in the \bp\ circumstellar gas.}

   \offprints{A.VM. (\email{alfred@iap.fr})}

   \institute{
   CNRS, UMR 7095, 
   Institut d'Astrophysique de Paris, 
   98$^{\rm bis}$ boulevard Arago, F-75014 Paris, France
   \and
   UPMC Univ. Paris 6, UMR 7095, 
   Institut d'Astrophysique de Paris, 
   98$^{\rm bis}$ boulevard Arago, F-75014 Paris, France
   \and
   School of Physics and Astronomy, 
   Tel-Aviv University, 
   Tel-Aviv 69978, Israel
   \and
   Observatoire de l'Universit\'e de Gen\`eve, 51 chemin des Maillettes, 1290, Sauverny, Switzerland
   }
   
   \date{} 
 
  \abstract{
$\text{The young planetary system }\beta$\,Pictoris is surrounded by a circumstellar disk of dust and gas. 
Because both dust and gas have a lifetime shorter than the system age, they need to be replenished continuously. 
The gas composition is partly known, but its location and its origin are still a puzzle. The gas source could be 
the exocomets (or so-called falling and evaporating bodies, FEBs), which are observed as transient features in 
absorption lines of refractory elements (Mg, Ca, and Fe) when they transit in front of the star at several tens of stellar radii. 
   
Nearly 1700~high-resolution spectra of $\beta$\,Pictoris have been obtained from 2003 to 2015 using the HARPS spectrograph. 
In these spectra, the circumstellar disk is always detected as a stable component among the numerous variable absorption signatures of transiting exocomets.

Summing all the 1700 spectra allowed us to reach a signal-to-noise ratio higher than 1000, which is an unprecedentedly high number for a \bp\ spectrum. It revealed many weak Fe I absorption lines of the circumstellar gas in more than ten excited states. 
These weak lines bring new information on the physical properties of the neutral iron gas in the circumstellar disk.
The population of the first excited levels follows a Boltzmann distribution with a slope consistent with a gas temperature 
of about 1300~K; this temperature corresponds to a distance to the star of $\sim$~38~R$_{\rm Star}$ and implies a turbulence 
of $\xi \sim$~0.8~km/s.

   
%
}


   \maketitle
%

\section{Introduction}
\label{Introduction}

When its IR excess was detected by the IRAS satellite in 1983, \bp\ (HD 39060; HR 2020) was the first star 
with a detection of a circumstellar disk that is seen edge-on (Smith \&\ Terrile 1984). 
Direct or indirect evidence of dust, gas, and falling and orbiting evaporating bodies (FEBs and OEBs, see, {\it e.g.,} 
Lecavelier et al.\ 1996a; Lecavelier 1998) and even planets were obtained very early 
(see review by Vidal-Madjar et al.\ 1998).

Absorption spectroscopic studies have revealed a stable gas component at the stellar radial velocity 
($\sim$~20~km/s in the heliocentric reference frame) and variable absorptions that have been attributed to transiting exocomets. 
The gas composition is relatively well constrained (Roberge et al.\ 2000). 
Emission line imaging also confirmed the presence of gas in the form of mainly Na\,I, Ca\,II, and Fe\,I (Brandeker et al.\ 2004). 

All these spectroscopic studies concluded that although the gas should be ejected away from the system by radiation pressure, 
it is observed in Keplerian rotation. Lagrange et al.\ (1998) proposed that atomic hydrogen could act as a braking gas at about 0.5~AU. Brandeker et al. (2004) observed different species in emission, and in particular, Ca\,II, Na\,I, and Fe\,I, which confirmed the braking gas scenario. Fern\'andez et al. (2006) later proposed that atomic carbon (mostly ionized) could 
be the required braking gas. C\,II, which is overabundant by more than ten times solar (Roberge et al.\ 2006),
could very efficiently brake all species via Coulomb interaction
simply because carbon suffers very little radiation pressure. 
This was also confirmed more recently by Brandeker (2011), who observed the Fe\,I and Na\,I lines in absorption and showed 
that the shift between the two sets of lines could also be explained by an overabundant C\,II braking gas.

However, the source of all the observed species within the so-called stable gas disk has yet to be found. 
It is clear that the FEBs are expelling gas, but it is not obvious that the gas observed in emission in the disk ({\it e.g.,} Brandeker et~al.\ 2004) necessarily comes from the FEBs. In particular, the C gas may be coming from dissociated CO released far from the star (Vidal-Madjar et al.\ 1994; Jolly et al.\ 1998). From the CO evaluated temperature, $\sim$20~K, a distance of about 
100~AU was estimated, allowing Lecavelier et al. (1996, 1998) to hypothesize that a second source could be related to distant OEBs. This was clearly confirmed by the recent CO detection with ALMA (Dent et al.\ 2014; Matr\`a et al.\ 2017) at about 80~AU from the star. All species origins could thus also be related to the accretion from this  
80~AU ``birth ring''. However, metals and C could have different origins.
If for metals the gas source is close to the central star, radiation pressure could push the gas at large distances where it is also observed. In this frame, the exocomets (or FEBs) 
could also be the source of the stable metal species that are
present in the circumstellar disk. The nature and properties of these infalling comets were constrained by observations and models 
(Ferlet et al. 1987; Beust et al. 1989, 1990; Beust \&\ Valiron 2007). 
More recently, detailed studies showed that these exocomets are distributed into two dynamical families 
(Kiefer et al. 2014). 

Our understanding of detailed mechanisms acting now in the young \bp\ planetary system could help us to reconstruct the puzzle of early phases in the young solar system. The late heavy bombardment, for instance, could have played an important role in the evolution of the Earth atmosphere.
Interaction with the central star, nearby interstellar clouds, supernova events, and cometary bombardments have built up an evolved Earth atmosphere and brought our oceans, where finally life appeared (see, {\it e.g. Central star:} Sekiya, Nakazawa \&\ Hayashi 1980; Watson, Donahue \&\ Walker 1981; Kasting \&\ Pollack 1983; Hunten, Pepin \&\ Walker 1987. {\it Interstellar clouds:} Hoyle \&\ Lyttleton 1939; McCrea 1975; Begelman \&\ Rees 1976; Talbot \&\ Newman 1977; Vidal-Madjar et al. 1978; Lallement et al. 2003; Gry \&\ Jenkins 2014. {\it Supernovae:} Rubenstein et al 1983; Trimble 1983; Li 2008. {\it Comets :} Rampino \&\ Stothers 1984; Morbidelli et al. 2000; Teiser \&\ Wurm 2009. {\it Cosmic-rays :} Laviolette 1987. {\it System formation:} Zahnle, Kasting \&\ Pollack 1988; Pepin 1991; Raymond, Quinn \&\ Lunine 2004; Genda \&\ Ikoma 2007; Kral et al. 2016. {\it Life:} Pavlov \&\ Klabunovskii 2015). 

More than 1700~high-resolution spectra of \bp\ were obtained with the HARPS spectrograph.
Stacking all these spectra into one, the total signal-to-noise
ratio (S/N) was higher than 1000 at shorter wavelengths 
and even 2000 at longer wavelengths. 
This high-resolution high-S/N spectrum revealed numerous newly detected weak absorption lines of the \bp\ circumstellar Fe\,I gas. 

Most importantly, the majority of these absorption features correspond to transition lines from initial excited states 
of neutral iron. Characterizing the distribution of the states of circumstellar Fe\,I can help to constrain 
the physical properties of the Fe\,I gas in the disk of \bp, under the influence of the stellar radiation.

In this paper we first present the observations (Sect.~\ref{Observations}). The measurements of the physical properties of the gas for various excitation levels of FeI are given in Sect.~\ref{stable_gas}. The results are discussed in Sect.~\ref{Discussion}. 

\begin{figure}
\begin{minipage}[b]{\columnwidth}         
\includegraphics[angle=90,trim=0.cm 0.cm 0.0cm 0.5cm,angle=0,clip=true,width=0.95\columnwidth]{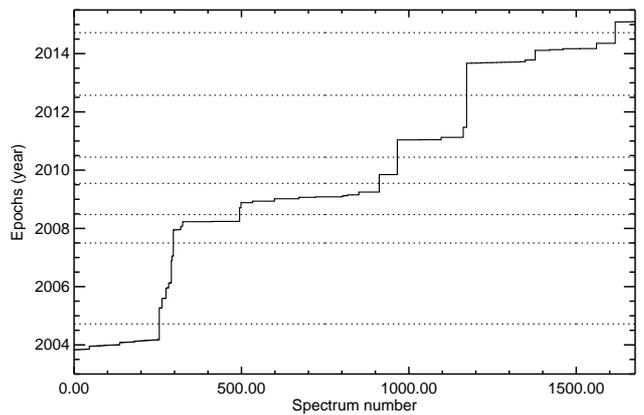}    
\hspace{0.36\textwidth}\\
\end{minipage}
\caption[]{
Time distribution of the HARPS observing nights used in our study. 
The horizontal dotted lines show the limits of various observation campaigns.
}
\label{repartition}
\end{figure}

\begin{figure*}
\begin{minipage}[b]{\textwidth}   
\includegraphics[trim=0.cm 0.cm 0.0cm 0.75cm,angle=90,clip=true,width=0.495\textwidth]{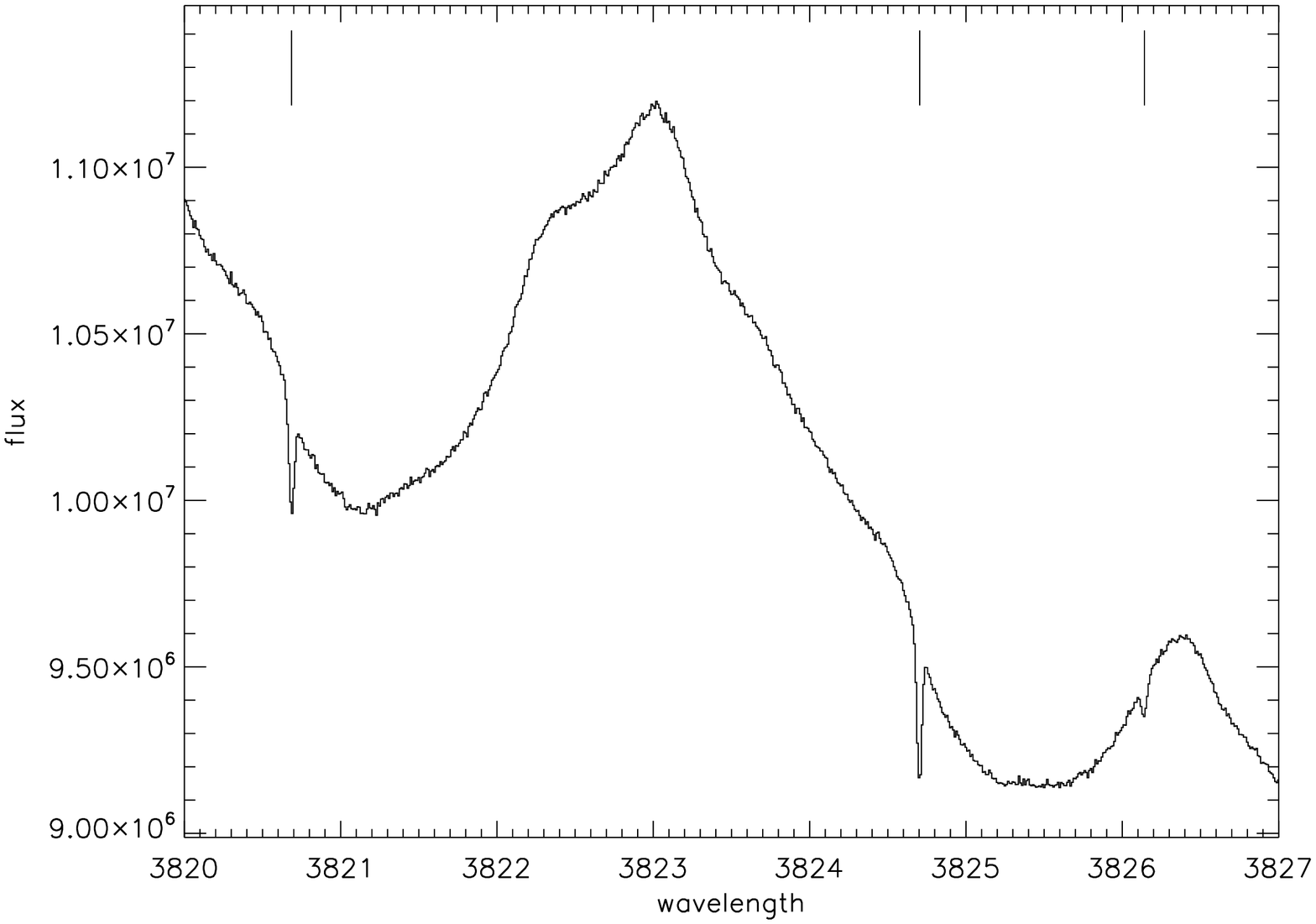}  
\includegraphics[trim=0.cm 0.cm 0.cm 0.75cm,angle=90,clip=true,width=0.495\textwidth]{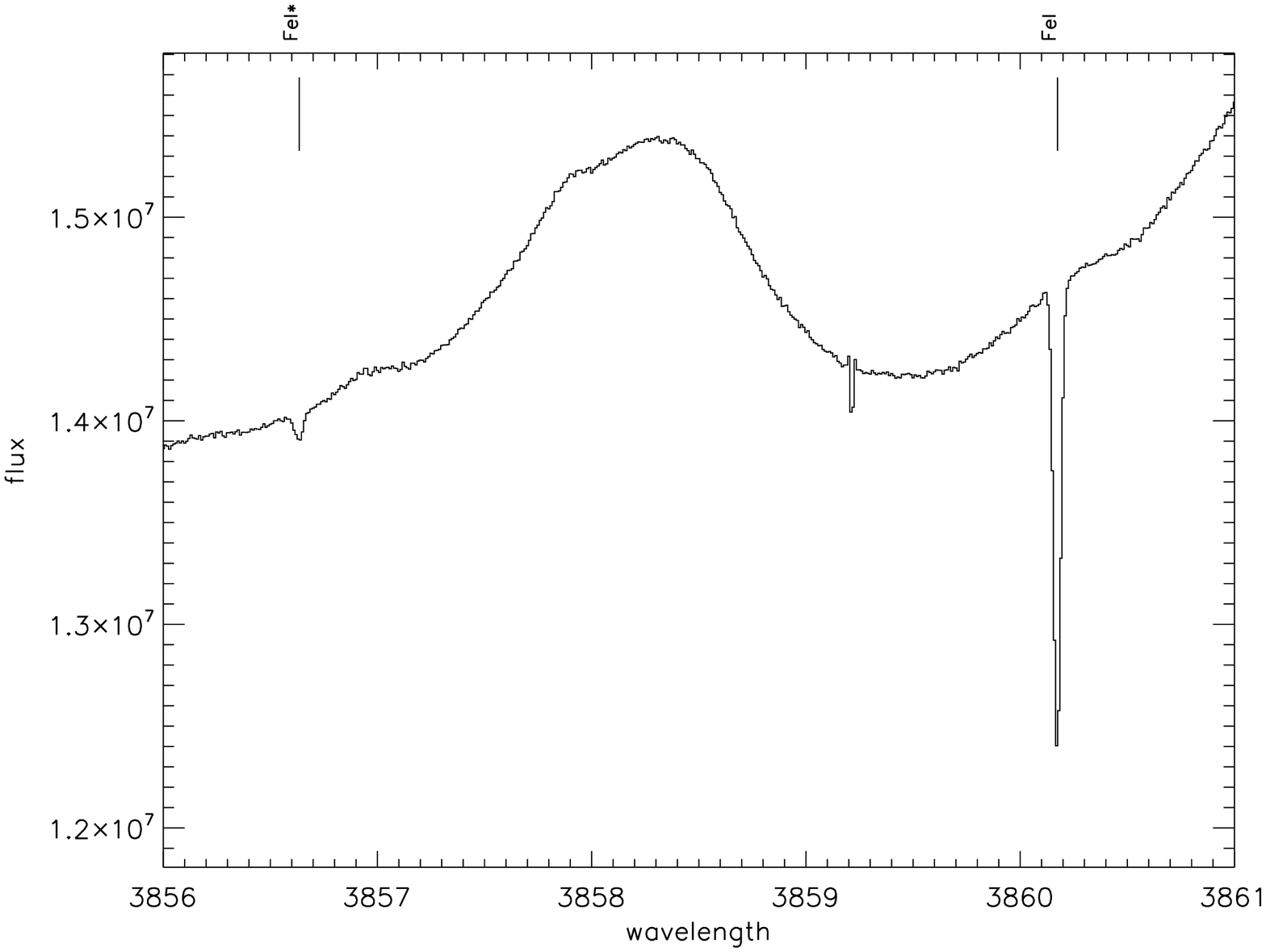}    
\hspace{0.36\textwidth}\\
\end{minipage}
\caption[]{
Averaged HARPS spectra of \bp\ (flux as a function of wavelength in \AA ). 
They reveal the very high S/N ratio achieved, even at very short wavelengths at~3820~\AA\ and 3860~\AA. 
Each detected Fe\,I line is indicated by a tick mark at +20.5 km/s, {\it i.e.,} the \bp\ heliocentric radial velocity. The detected signatures are due to the stable gas within the circumstellar disk. 

{\bf Left:} Section of the spectrum from 3820 to 3827~\AA . Three lines are detected: one line from the 6928\,cm$^{-1}$ excited 
levels at 3820.7~\AA , one line from the ground base level near 3824.7~\AA , and one line from the 7377\,cm$^{-1}$ excited levels at 3823.15~\AA . 

{\bf Right:} Spectral region from 3856 to 3861~\AA. The strongest Fe\,I line from ground base level is seen near 3860.15~\AA, 
as well as the line from the 416\,cm$^{-1}$ excited level (noted FeI* near 3856.65~\AA ). 
}
\label{Fig:Observations}
\end{figure*}

\section{Observations}
\label{Observations}

The \bp\ syxtem has been monitored with the HARPS spectrograph mounted on the 3.6m telescope of La Silla (ESO Chile) 
from 2003 to 2015 on a (mostly) regular basis (except between 2004 and 2007 and in 2012).
The spectra were primarily obtained to constrain the mass of planets around 
\bp\ using radial velocity measurements 
(Lagrange et al.\ 2012) and to analyze the statistics of the exocomets (Kiefer et al. 2014).
Since \bp\ is observable only during the summer in the southern hemisphere, the observations were carried out 
essentially from September to April of each season. The 1D spectra were extracted via the standard most recent 
HARPS pipeline (DRS 3.5), including localization of the spectral orders in the 2D images, optimal order extraction, 
cosmic-ray rejection, wavelength calibration, flat-field corrections, and 1D reconnection of the spectral orders 
after correction for the blaze. We organized the spectra into different samples, each constituting one summer 
of observations, with the exception of the period 2004-2007. During this period, the average number of spectra observed 
per summer is 14, which is very small compared to the other period averages (234 spectra per summer). 
For this reason, we created a special sample covering the three summers from 2004 to 2007. 
This repartition is summarized in Table~\ref{tab:repartition} and Fig.~\ref{repartition}. 

\begin{table}
\caption{Repartition of HARPS observations. The S/N is calculated next to the main Fe I line between 3859.30\,\AA~and 3859.70\,\AA.}
\label{tab:repartition}
\begin{tabular}{crcc}
\hline
Periods    &   $N_\text{obs}$  & Total S/N &  Average Epoch  \\ 
 (year)    &                   & at 3860\AA &    (year)      \\
\hline
2003-2004  & 255               &  962      &    2004.025 \\
2004-2007  & 42                &  390      &    2006.235 \\
2007-2008  & 198               &  492      &    2008.203 \\
2008-2009  & 417               &  909      &    2009.070 \\
2009-2010  & 54                &  278      &    2009.847 \\
2010-2011  & 207               &  467      &    2011.081 \\
2013-2014  & 453               &  799      &    2013.929 \\
2014-2015  & 60                &  493      &    2015.091 \\
\hline
\end{tabular}
\end{table}

The 1D spectra of each period are all summed, weighting each spectrum $F_i$ with its square 
of the {$\text{(S/N)}_i$} at order 33 (with a median wavelength of 4745.55\,\AA).
This order was selected because of its proximity to the median HARPS order 
while being as flat as possible and not affected by any broad stellar line or telluric signature. 
The following formula was used to derive the summed spectra:
\begin{equation}
F_\text{sum} = \frac{\sum_{i=1}^{N_\text{spec}} F_i \times \text{(S/N)}_i^2}{\sum_{i=1}^{N_\text{spec}} \text{(S/N)}_i^2},
\end{equation}
where $N_\text{spec}$ is the total number of spectra $F_i$, and {$\text{(S/N)}_i$} is the 
signal-to-noise ratio of each spectrum~$i$. 

In the spectra coaddition, no resampling was needed thanks to the high stability of the spectrograph 
and the HARPS data reduction software (DRS), which produces the spectra on the same wavelength scale for all spectra 
in the solar system barycentric reference frame \footnote{eso.org/sci/facilities/lasilla/instruments/harps/doc/DRS.pdf}. 
We checked that the zero flux level of the 1D spectra was always well defined. To this purpose, we
considered the tip of the Ca II circumstellar (CS) absorption line, which should be zero. 
We found that 122 of 1686~spectra have a negative flux at the tip of the Ca II CS line
at about -10~ADU, 
which is negligible with respect to the typical flux 
around the Fe\,I lines, of about 10$^4$~ADU.
Moreover, the S/Ns of these 122 spectra are among the lowest in the full sample.   

Coadding the spectra greatly improved the S/N of \bp\ spectrum in each period. 
We give in Table~\ref{tab:repartition} the value of the S/N around the main Fe I CS line at 3860\,\AA. 
We calculated it by fitting the flux continuum by a second-degree polynomial between 3859.30\,\AA~and 3859.70\,\AA, 
a rather flat region next to the Fe I CS line. 
We estimated the noise from the standard deviation of the residuals and the signal from the average flux. 

Because the stable gas is continuously present in the \bp\ spectra, all possible remaining variable spectral 
features are washed away in the stacking process. We ran the procedure defined in the Kiefer et al.~(2014)
to recover the \bp\ reference spectrum with no variable absorptions 
by searching for the highest flux over each pixel at each wavelength. 
This approach was necessary in the case of the CaII lines, where almost every spectrum includes an exocomet transit event. 
Here we checked the possible pollution of the strongest Fe\,I absorption lines by exocomet transits
using this procedure. We found that the line profiles obtained using this process 
were identical to the profile obtained using a simple coaddition process, 
showing that the pollution by FEB signatures is negligible.
A few FEBs were observed in Fe\,I (Welsh \&\ Montgomery 2015, 2016); there is, however, a very low number of detections, and all of them are found to be at high radial velocity, away from the stable gas absorption signature, which remains at the stellar radial velocity. 
The pollution of the Fe\,I profile by transiting exocomets is therefore negligible. 

Finally, we stacked all gathered spectra and averaged them in order to improve the S/N. Given the high stability of the spectrograph,  the spectral resolution is not degraded (R $\sim$ 115000; Mayor
et al. 2003) after the stacking and averaging process. This resolving power was furthermore independently confirmed by Brandeker (2014), who found that in HARPS observations common to ours, the resolving power was on the order of 112000.
The result is spectacular, as shown in Figure~\ref{Fig:Observations}. The S/N reached is on the order of 1000 at the shorter wavelengths and up to 2000 at longer wavelengths of the HARPS spectrum.

\section{Stable gas in \bp\ }
\label{stable_gas}
  
\begin{table*}
\centering
\caption{FeI lines (i lower level, k upper level). 
The equivalent widths given in the last column are derived from the column densities measured in 
Sect.\ref{FeI_lines},   which are listed in Table~\ref{FeI_levels_N}.
}
\begin{tabular}{cccccc}
 \hline
 FeI lines   & A$_{\rm ki}$  & f$_{\rm ik}$ & E$_{\rm i}$ & E$_{\rm k}$ & W$_{\rm eq}$\\
  (\AA ) & (s$^{-1}$) &  & (cm$^{-1}$) & (cm$^{-1}$)& (m\AA ) \\ 
 \hline
3795.002 & 1.15$\times$10$^7$ & 3.47$\times$10$^{-2}$ & 7986 & 34329 & 0.047 \\
3799.547 & 7.31$\times$10$^6$ & 2.04$\times$10$^{-2}$ & 7728 & 34040 & 0.019 \\
3812.964 & 7.91$\times$10$^6$ & 1.23$\times$10$^{-2}$ & 7728 & 33947 & 0.011 \\
3815.840 & 1.12$\times$10$^8$ & 1.90$\times$10$^{-1}$ & 11976 & 38175& 0.115  \\ 
3820.425 & 6.67$\times$10$^7$ & 1.20$\times$10$^{-1}$ & 6928 & 33096 & 1.269 \\
3824.443 & 2.83$\times$10$^6$ & 4.83$\times$10$^{-3}$ &    0 & 26140 & 1.569 \\
3825.881 & 5.97$\times$10$^7$ & 1.02$\times$10$^{-1}$ & 7377 & 33507 & 0.400 \\
3827.822 & 1.05$\times$10$^8$ & 1.65$\times$10$^{-1}$ & 12561 & 38678& 0.044 \\
3834.222 & 4.52$\times$10$^7$ & 7.13$\times$10$^{-2}$ & 7728 & 33802 & 0.067 \\
3840.437 & 4.70$\times$10$^7$ & 6.24$\times$10$^{-2}$ & 7986 & 34017 &0.086 \\
3841.047 & 1.36$\times$10$^8$ & 1.80$\times$10$^{-1}$ & 12969 & 38996& 0.030  \\ 
3849.966 & 6.05$\times$10$^7$ & 4.49$\times$10$^{-2}$ & 8155 & 34122 & 0.027 \\
3856.371 & 4.64$\times$10$^6$ & 7.39$\times$10$^{-3}$ &  416 & 26340 & 0.382 \\
3859.911 & 9.69$\times$10$^6$ & 2.17$\times$10$^{-2}$ &    0 & 25900 & 7.182 \\
3865.523 & 1.55$\times$10$^7$ & 3.47$\times$10$^{-2}$ & 8155 & 34017 & 0.021 \\
3878.573 & 6.17$\times$10$^6$ & 8.36$\times$10$^{-3}$ &  704 & 26479 & 0.245 \\
3886.282 & 5.29$\times$10$^6$ & 1.20$\times$10$^{-2}$ &  416 & 26140 & 0.630 \\
3895.656 & 9.39$\times$10$^6$ & 7.13$\times$10$^{-3}$ &  888 & 26550 & 0.082 \\
3899.707 & 2.58$\times$10$^6$ & 5.89$\times$10$^{-3}$ &  704 & 26340 & 0.174 \\
3906.479 & 8.32$\times$10$^5$ & 1.90$\times$10$^{-3}$ &  888 & 26479 & 0.022 \\
3920.257 & 2.60$\times$10$^6$ & 1.79$\times$10$^{-2}$ &  978 & 26479 & 0.072 \\
3922.911 & 1.08$\times$10$^6$ & 3.19$\times$10$^{-3}$ &  416 & 25900 & 0.171 \\
3927.919 & 2.60$\times$10$^6$ & 1.00$\times$10$^{-2}$ &  888 & 26340 & 0.117  \\
3930.296 & 1.99$\times$10$^6$ & 6.46$\times$10$^{-3}$ &  704 & 26140 & 0.194 \\
4045.812 & 8.62$\times$10$^7$ & 2.12$\times$10$^{-1}$ & 11976 & 36686& 0.144 \\
4063.594 & 6.65$\times$10$^7$ & 1.65$\times$10$^{-1}$ & 12561 & 37163& 0.049  \\
4071.738 & 7.64$\times$10$^7$ & 1.90$\times$10$^{-1}$ & 12969 & 37521 & 0.036 \\
4271.760 & 2.28$\times$10$^7$ & 7.62$\times$10$^{-2}$ & 11976 & 35379 & 0.058 \\ 
4307.902 & 3.38$\times$10$^7$ & 1.21$\times$10$^{-1}$ & 12561 & 35768 & 0.041 \\
4325.762 & 5.16$\times$10$^7$ & 2.03$\times$10$^{-1}$ & 12969 & 36079 & 0.043 \\
4383.544 & 5.00$\times$10$^7$ & 1.76$\times$10$^{-1}$ & 11976 & 34782 & 0.140 \\
4404.750 & 2.75$\times$10$^7$ & 1.03$\times$10$^{-1}$ & 12561 & 35257 & 0.036 \\
 \hline
\end{tabular}
\label{observations}
\end{table*}

Because of the very high S/N ratio achieved, we are able to detect numerous Fe\,I lines 
in the ground state as well as in several excited levels. 
All the detected Fe\,I lines are listed in Table~\ref{observations}.

\begin{table}
\centering
\caption{Fe\,I excitation levels corresponding to the detected absorption lines. The lower levels of the 12 excited levels range from 416 to 12969\,cm$^{-1}$.}
\begin{tabular}{rrrr}
 \hline
 Energy level   & Energy level & J  & Number   \\
 (cm$^{-1}$)   &  (K)  &  & of lines  \\
 \hline
  0  &  0  & 4 & 2  \\
416  & 598 & 3 & 3  \\
704 & 1013 & 2 & 3  \\
888 & 1278 & 1 & 3  \\
978 & 1407 & 0 & 1  \\
6928 & 9968 & 5 & 1  \\
7377 & 10614 & 4 & 1  \\
7728 & 11120 & 3 & 3  \\
7986 & 11490 & 2 & 2  \\
8155 & 11733 & 1 & 2  \\
11976 & 17232 & 4 & 4  \\
12561 & 18073 & 3 & 4  \\
12969 & 18660 & 2 & 3  \\
\hline
\end{tabular}
\label{levels}
\end{table}

The profiles of the the strongest lines of the 12 Fe\,I excited levels that we detected 
are shown in Fig.~\ref{Strongest_FeI_excited_fits}. 
Lines from excited levels up to 12969 cm$^{-1}$ are clearly detected. 
Several absorption lines of highly excited levels have a low column density, but they lie in spectral 
regions above 4000~\AA , where the high flux level yields a high S/N, which allows sensitive detections and 
accurate measurement of column densities. 

\begin{figure*}
\begin{minipage}[b]{\textwidth}   
\includegraphics[trim=0.cm 0.cm 0.cm 0.75cm,angle=90,clip=true,width=0.495\textwidth]{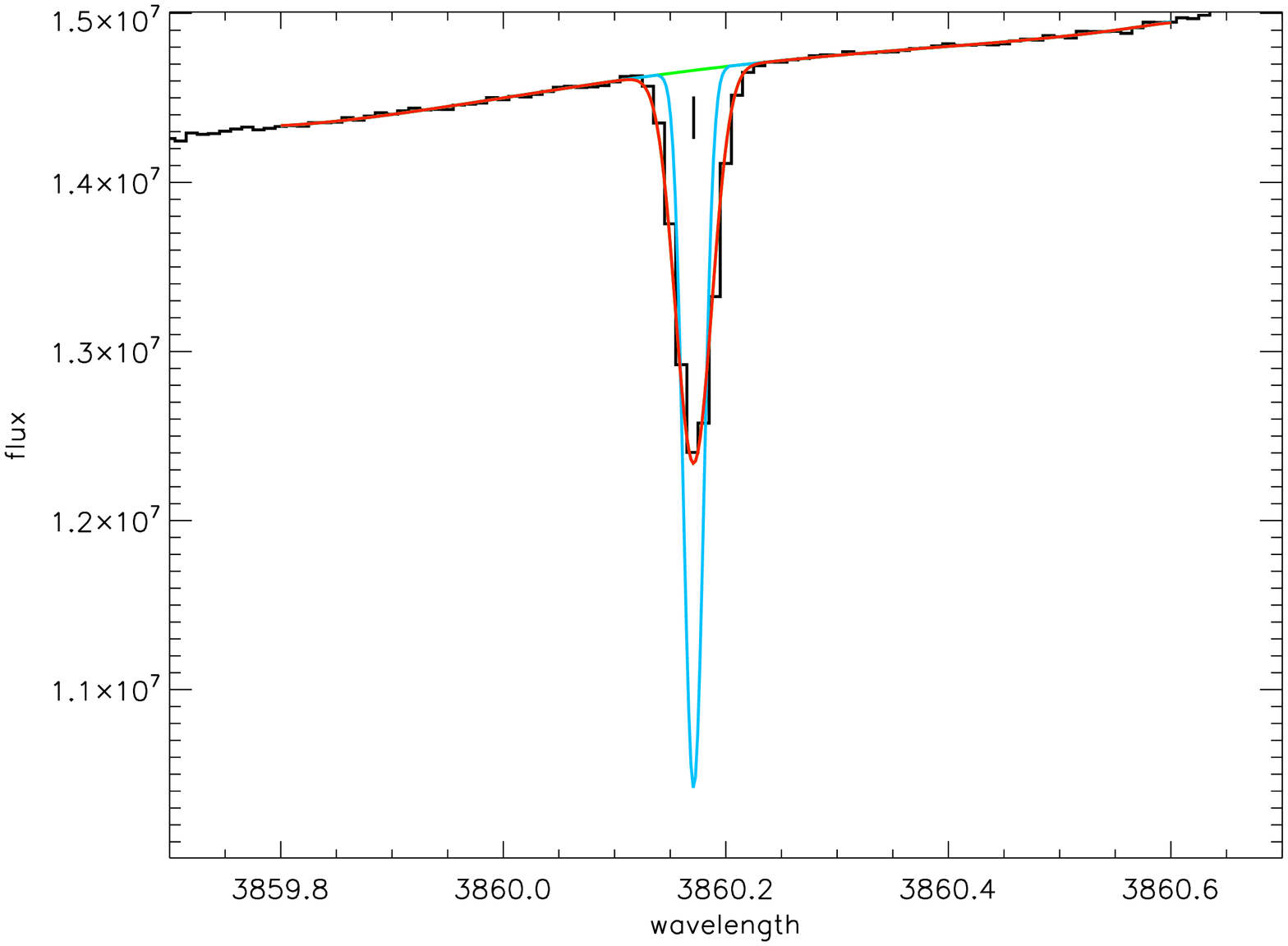}  
\includegraphics[trim=0.cm 0.cm 0.cm 0.75cm,angle=90,clip=true,width=0.495\textwidth]{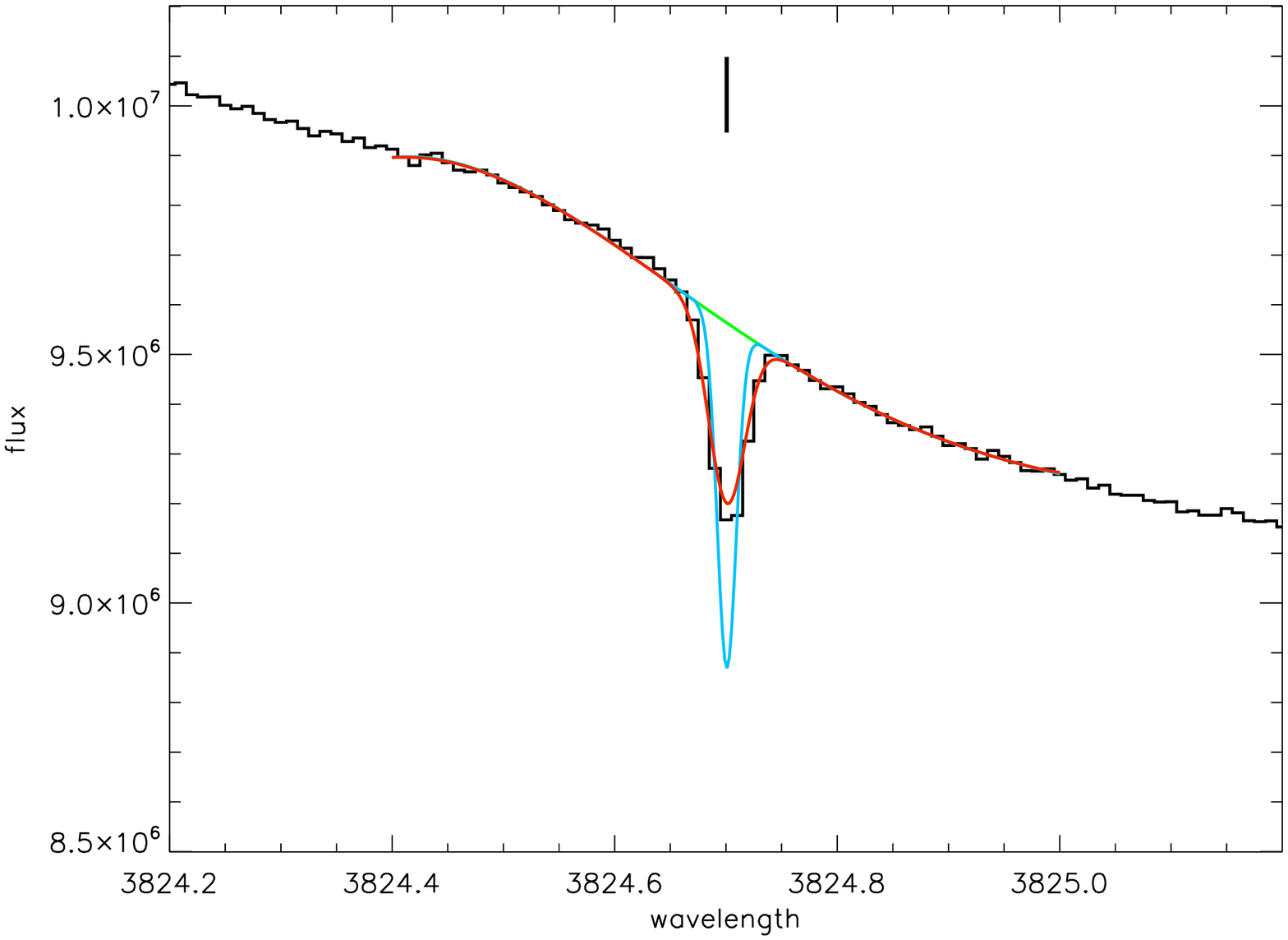}  
\hspace{0.18\textwidth}\\
\end{minipage}
\caption[]{Two Fe\,I \bp\ lines from the ground base level (flux in arbitrary units 
as a function of wavelength in \AA ). The observations are shown as black histograms, 
the fitted stellar continuum as green lines. 
The blue lines represent the intrinsic absorption feature, 
and the red lines are its convolution with the HARPS instrumental profile. 
In the two lines, the intrinsic absorption profiles (blue lines) are far from saturated, 
implying that the column density of Fe\,I from the ground base level can be well constrained.
}
\label{FeI_HARPS}
\end{figure*}

\subsection{Circumstellar gas in the ground base Fe\,I lines}
\label{FeI_lines}

We used the profile-fitting procedure
{\tt Owens.f}, developed by Lemoine M. and the FUSE French Team (see
{\it e.g.} Lemoine et al. 2002; H{\'e}brard et al. 2002) for
the analysis. 
The absorption lines were fit by Voigt profiles using $\chi ^2$ minimization.
Using {\tt Owens.f}, we searched for the best simultaneous fit to all detected spectral lines assuming
that all line profiles are the result of absorption by a low number of components with, 
for a given component, the same values for the radial velocity, $V$, the temperature, $T$, 
the turbulent broadening, $\xi$, and the column densities for each species {\bf s}, $N_{\bf s}$. 
The width of the lines, $b$, results from both parameters $T$ and $\xi$, with b$^2$=(2kT/m)$^2$+$\xi$\,\!$^2$ 
(where $k$ is the Boltzmann constant and $m$ is the mass of the considered species). 
The degeneracy in the measurement of $T$ and $\xi$ can be left only when several species with different masses 
are fit simultaneously, which is not the case here because we considered only the iron absorption features.
We emphasize that the column densities and their associated error bars were derived from a global fit 
of the lines profiles and not from the equivalent width measurements. The equivalent widths listed in 
Table~\ref{observations} are given for comparison purposes.

In a first step, a single component was considered to fit the two observed Fe\,I lines from the ground base level. 
The stellar continuum was modeled by a fourth-order polynomial, and the absorption profile by the convolution 
of the instrument line spread function (LSF) with a Voigt profile defined by its standard parameters ($V$, $b,$ and $N$). 
The fitted profiles are shown in Fig.~\ref{FeI_HARPS}, and the corresponding parameters 
are listed in Table~\ref{FeI_parameters}.

The line width ($b$) is on the order of 0.8~km/s, which is very narrow (the HARPS instrumental width is 2.6~km/s) 
and in contrast with all published $b$ values for the \bp\ stable circumstellar component 
measured using the complex Ca\,II doublet, which are all on the order of 2~km/s. 
We interpret this low value of $b$ as the combination of two factors: 
1) iron has a higher mass than calcium, leading to a lower line width in its thermal component, 
2) the Fe\,I line is less complex, with no stellar chromospheric emission variations and no significant 
pollution by variable exocomet absorption events. 
As a result, the line width measured here is likely closer to the
real physical value. 
This narrow line width furthermore shows that the stacking process of hundreds of HARPS spectra 
does not significantly affect the line profile 
(in agreement with the expected stability for the HARPS spectrograph). 

To evaluate the error bars on the parameter estimates, we calculated the $\Delta\chi^2$ variations 
resulting from the change of the tested parameters, while other parameters were let free to vary 
(see H\'ebrard et al.\ 2002). The resulting error bars for the Fe\,I radial velocity V$_{\rm FeI}$, 
line width b$_{\rm FeI}$ , and column density N$_{\rm FeI}$ are given in Table~\ref{FeI_parameters}. 

The estimated column density of Fe\,I is robust because the two lines are unsaturated and the resulting 
equivalent width is in the linear part of the curve of growth; the estimate is thus independent 
of the velocity distribution of the absorbing gas. This column density corresponds to an equivalent width 
of 7.1\,m\AA\ for the 3860\,\AA\ line, which is consistent with the equivalent width measured by
Welsh \& Montgomery (2016). The variations in column density detected by Welsh \& Montgomery (2016) 
do not affect the scientific conclusions derived here and will be discussed 
in a forthcoming paper (Kiefer et al., in preparation).

\begin{table}
\centering
\caption{Measurements for the Fe\,I ground base level obtained with a simultaneous fit of the profiles of the two lines.}
\begin{tabular}{ccc}
 \hline
  V$_{\rm FeI}$   & b$_{\rm FeI}$ & N$_{\rm FeI}$  \\
  Heliocentric radial   & Line width & Column density \\
  velocity (km/s) & (km/s)  & ($\times$10$^{13}$ cm$^{-2}$) \\
 \hline
    20.19$\pm 0.05$  & 0.80$\pm 0.07$ & 0.251($\pm 0.01)$       \\
 \hline
\end{tabular}
\label{FeI_parameters}
\end{table}

\begin{figure*}
\begin{minipage}[b]{\textwidth}   
\includegraphics[trim=0.cm 0.cm 0.cm 0.75cm,angle=90,clip=true,width=0.33\textwidth]{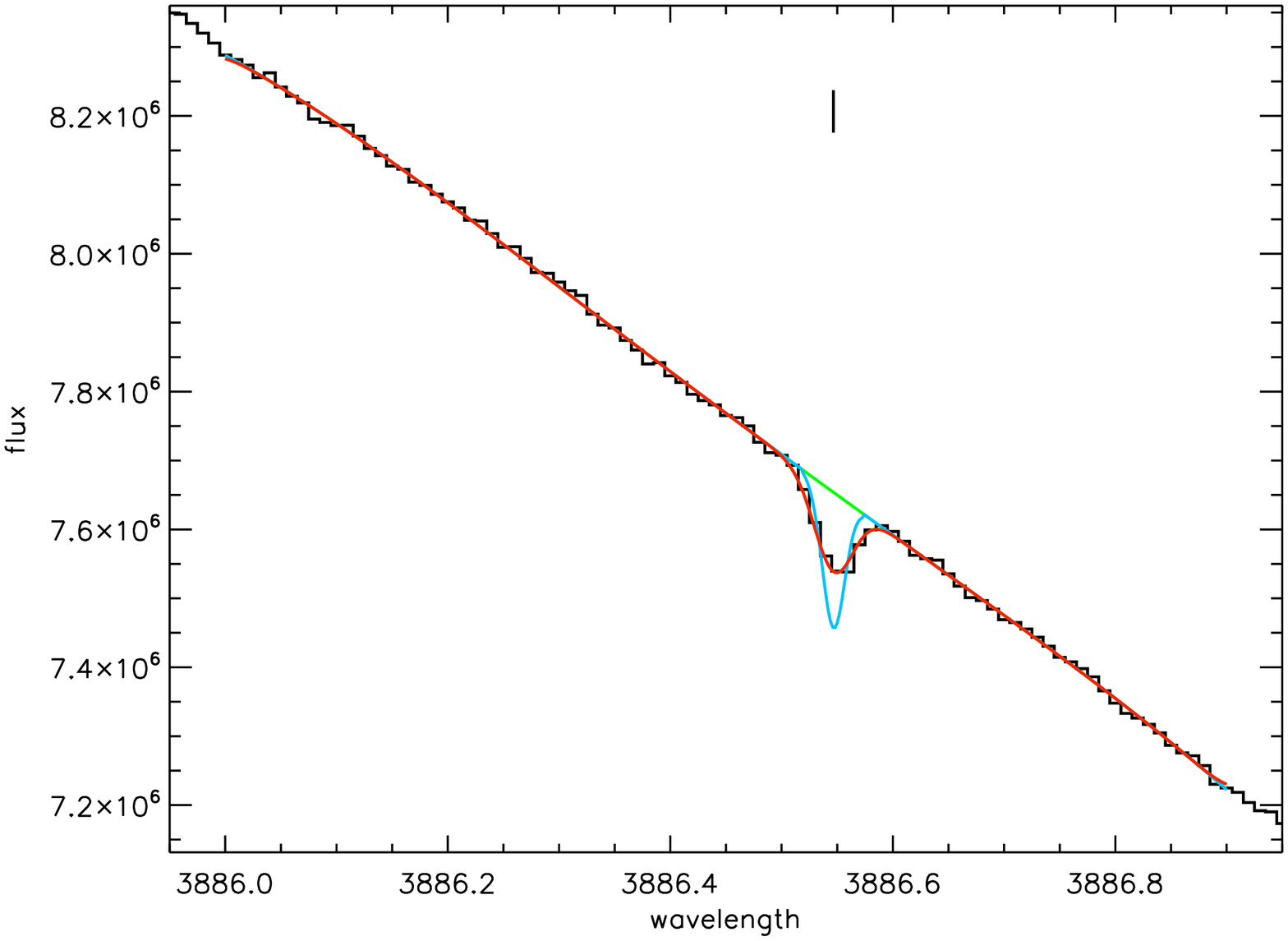}    
\includegraphics[trim=0.cm 0.cm 0.cm 0.75cm,angle=90,clip=true,width=0.33\textwidth]{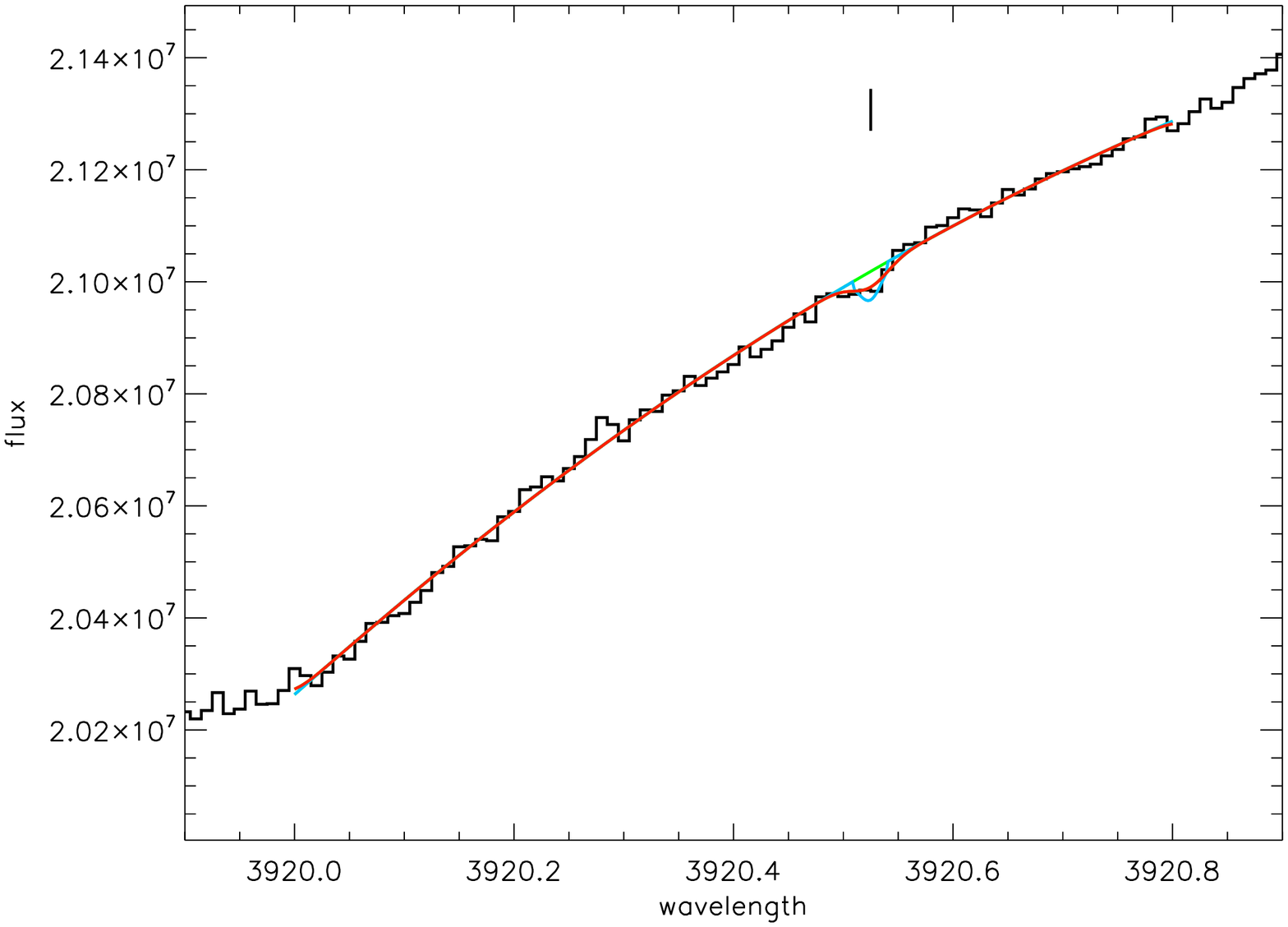}    
\includegraphics[trim=0.cm 0.cm 0.cm 0.75cm,angle=90,clip=true,width=0.33\textwidth]{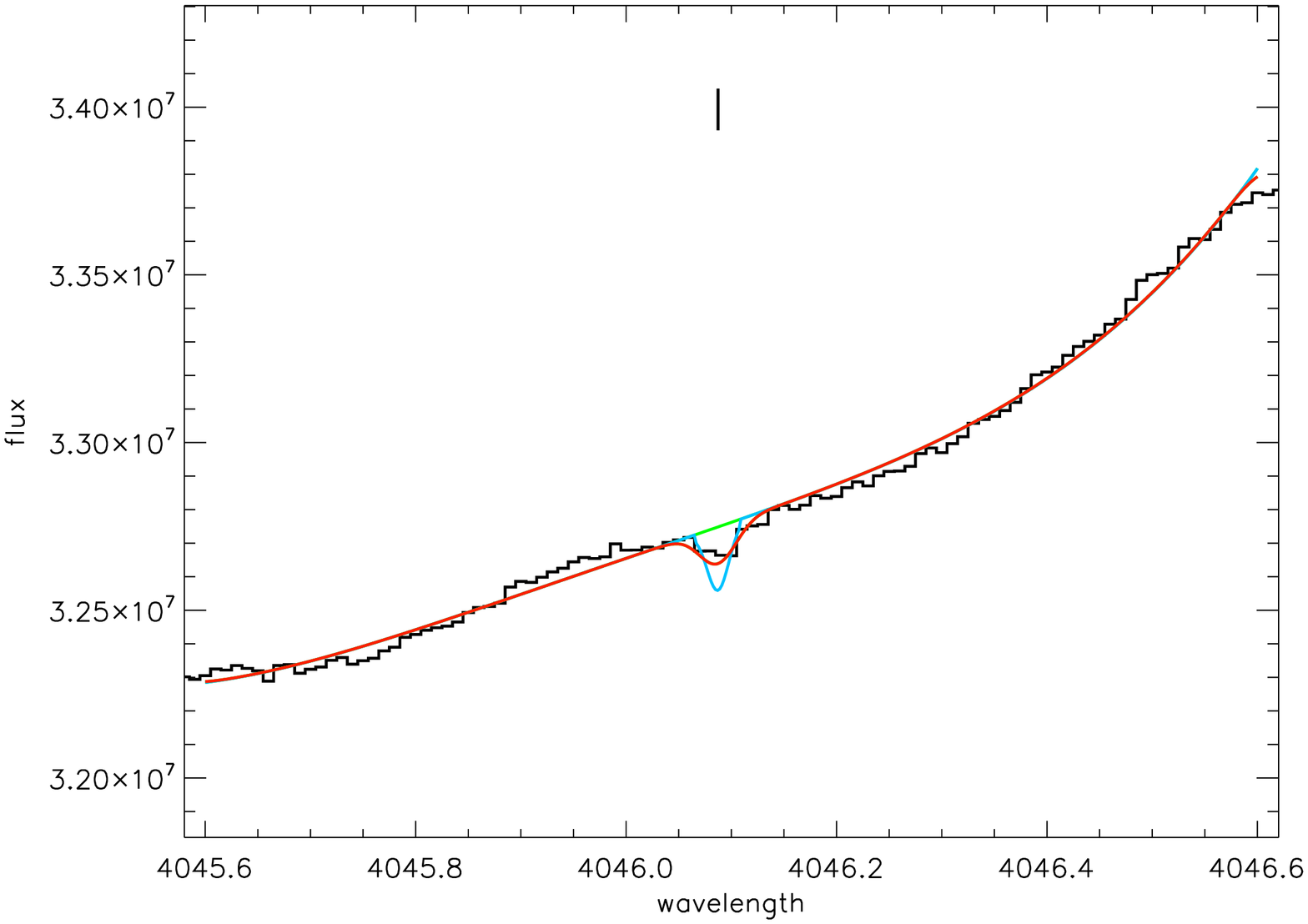}   
\hspace{0.72\textwidth}\\
\includegraphics[trim=0.cm 0.cm 0.cm 0.75cm,angle=90,clip=true,width=0.33\textwidth]{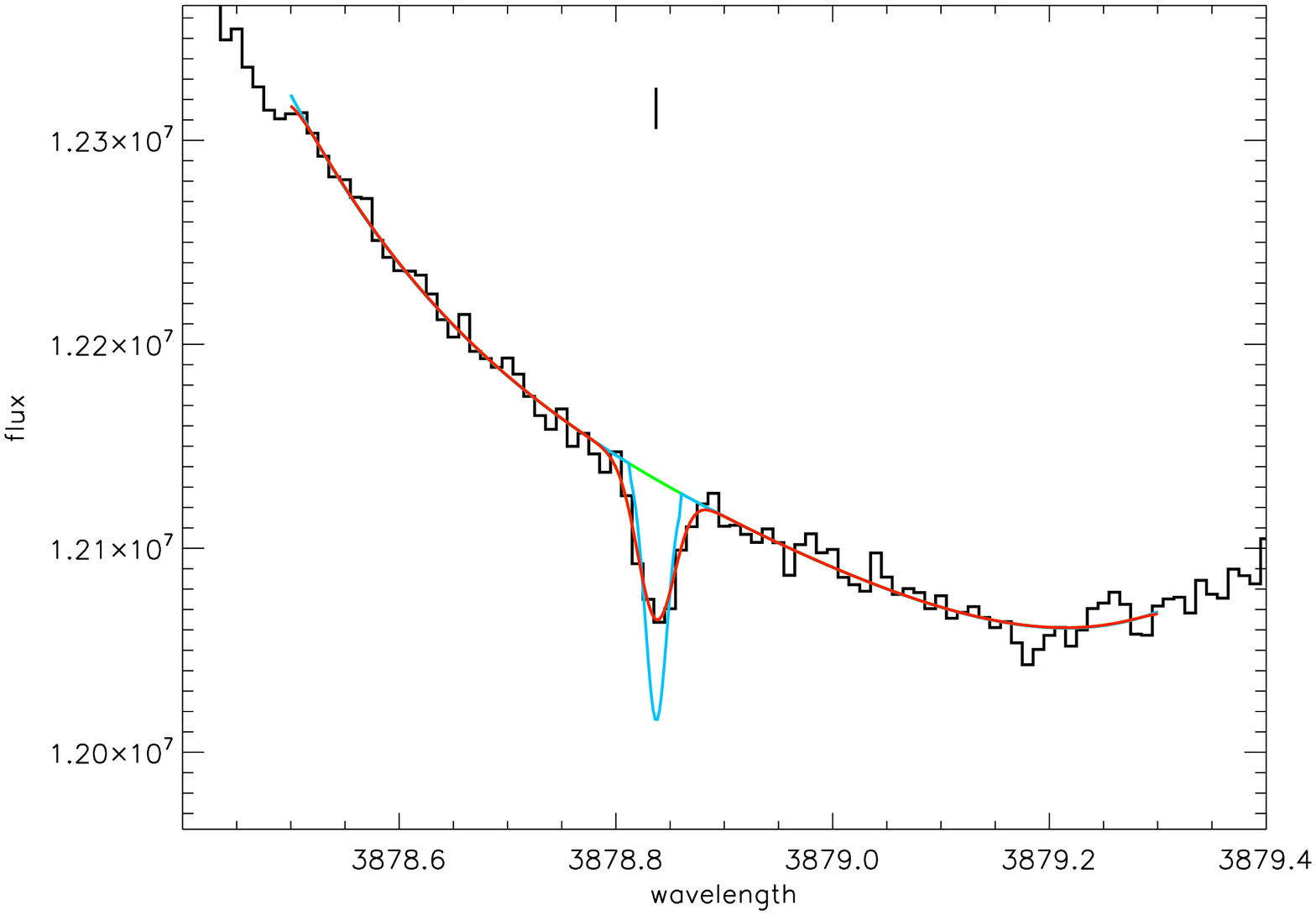}    
\includegraphics[trim=0.cm 0.cm 0.cm 0.75cm,angle=90,clip=true,width=0.33\textwidth]{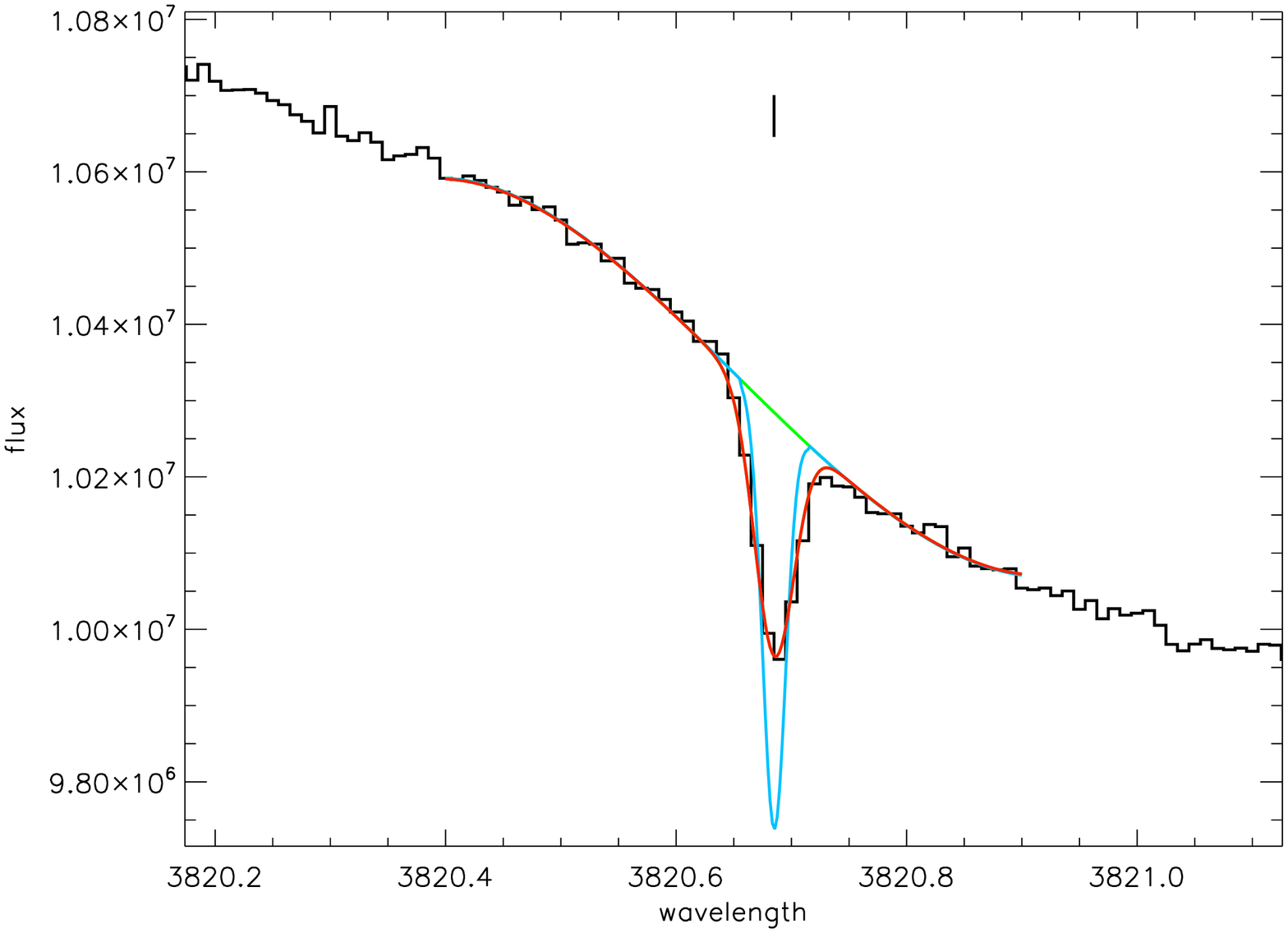}    
\includegraphics[trim=0.cm 0.cm 0.cm 0.75cm,angle=90,clip=true,width=0.33\textwidth]{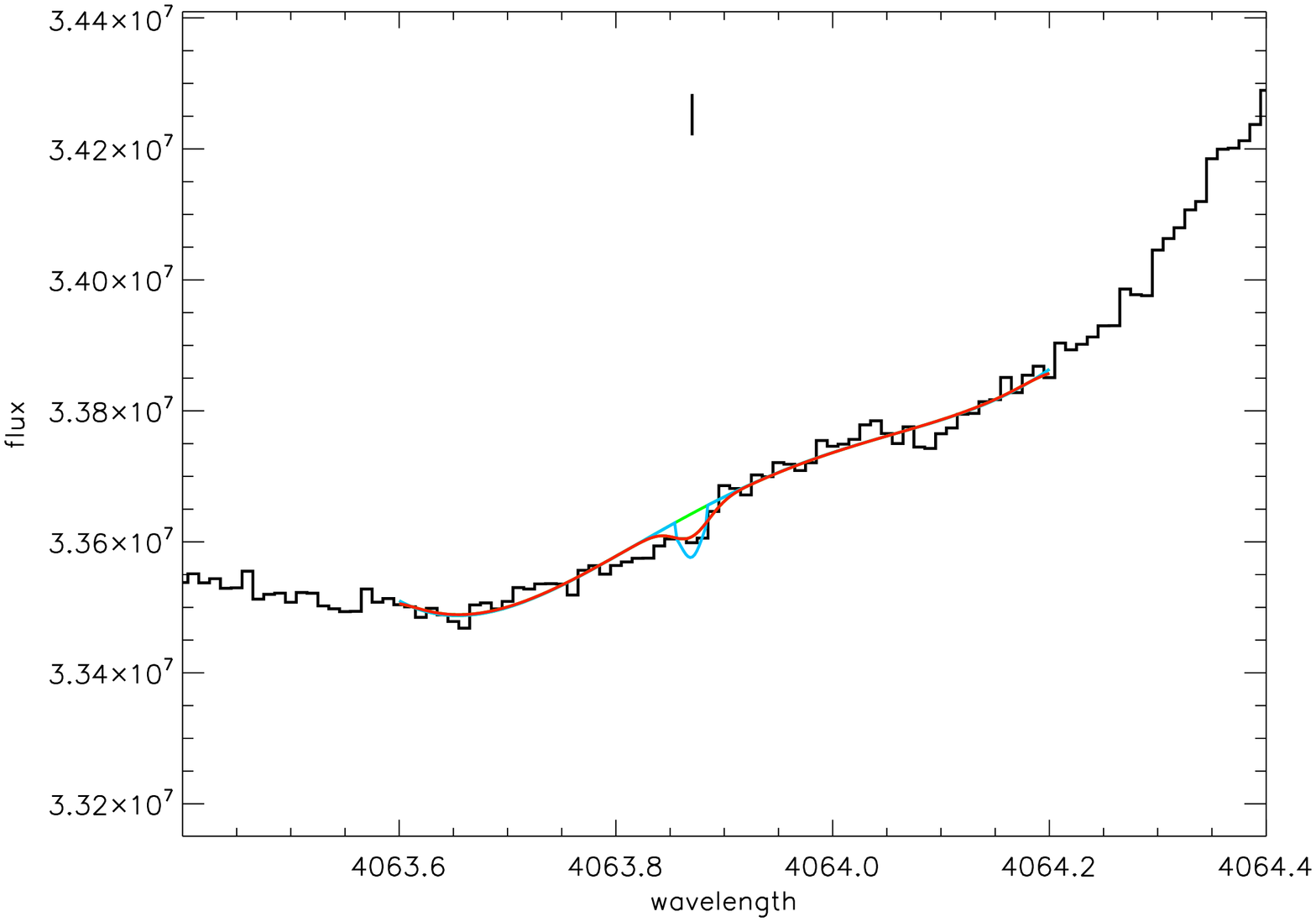}   
\hspace{0.72\textwidth}\\
\includegraphics[trim=0.cm 0.cm 0.cm 0.75cm,angle=90,clip=true,width=0.33\textwidth]{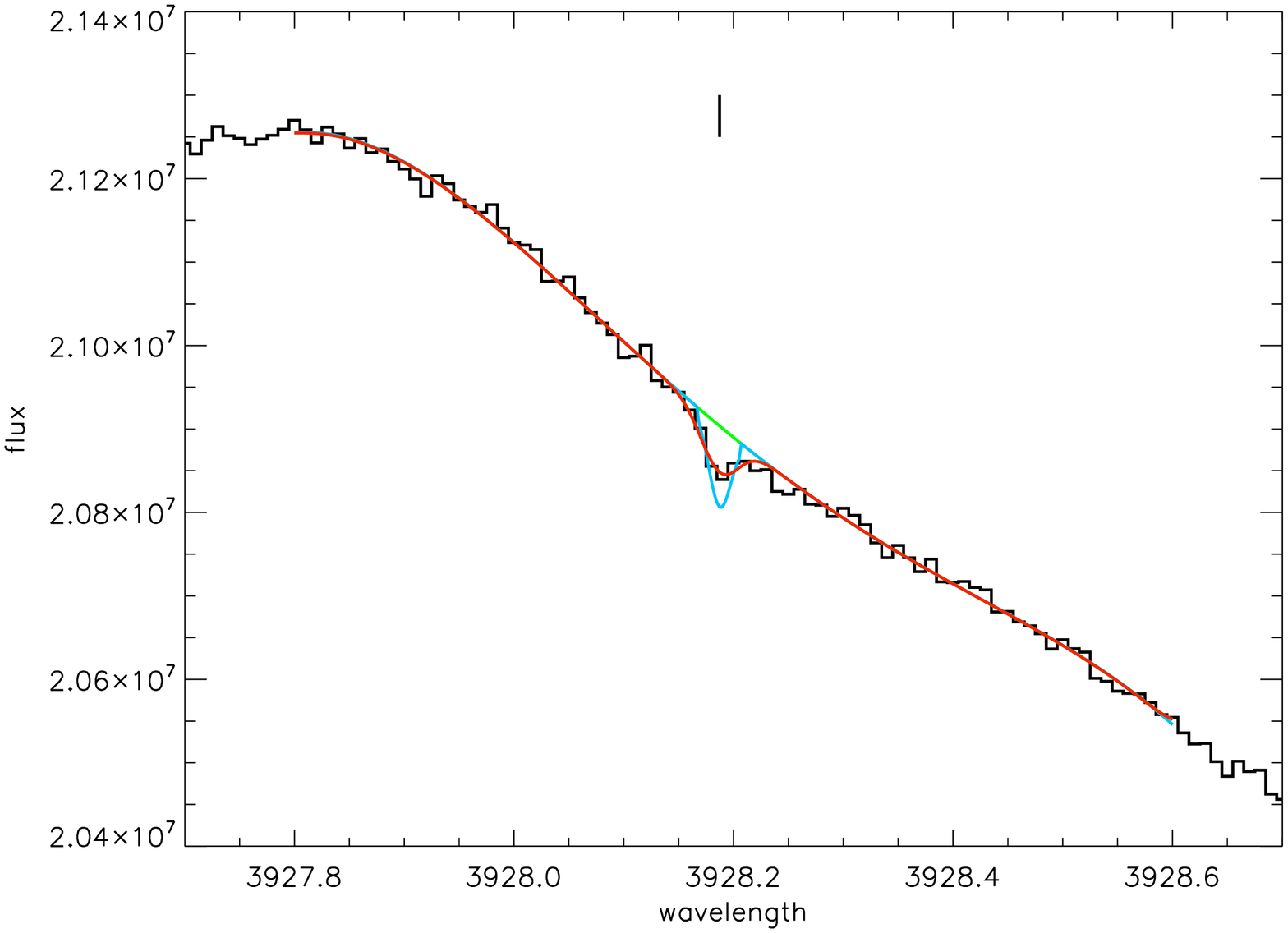}    
\includegraphics[trim=0.cm 0.cm 0.cm 0.75cm,angle=90,clip=true,width=0.33\textwidth]{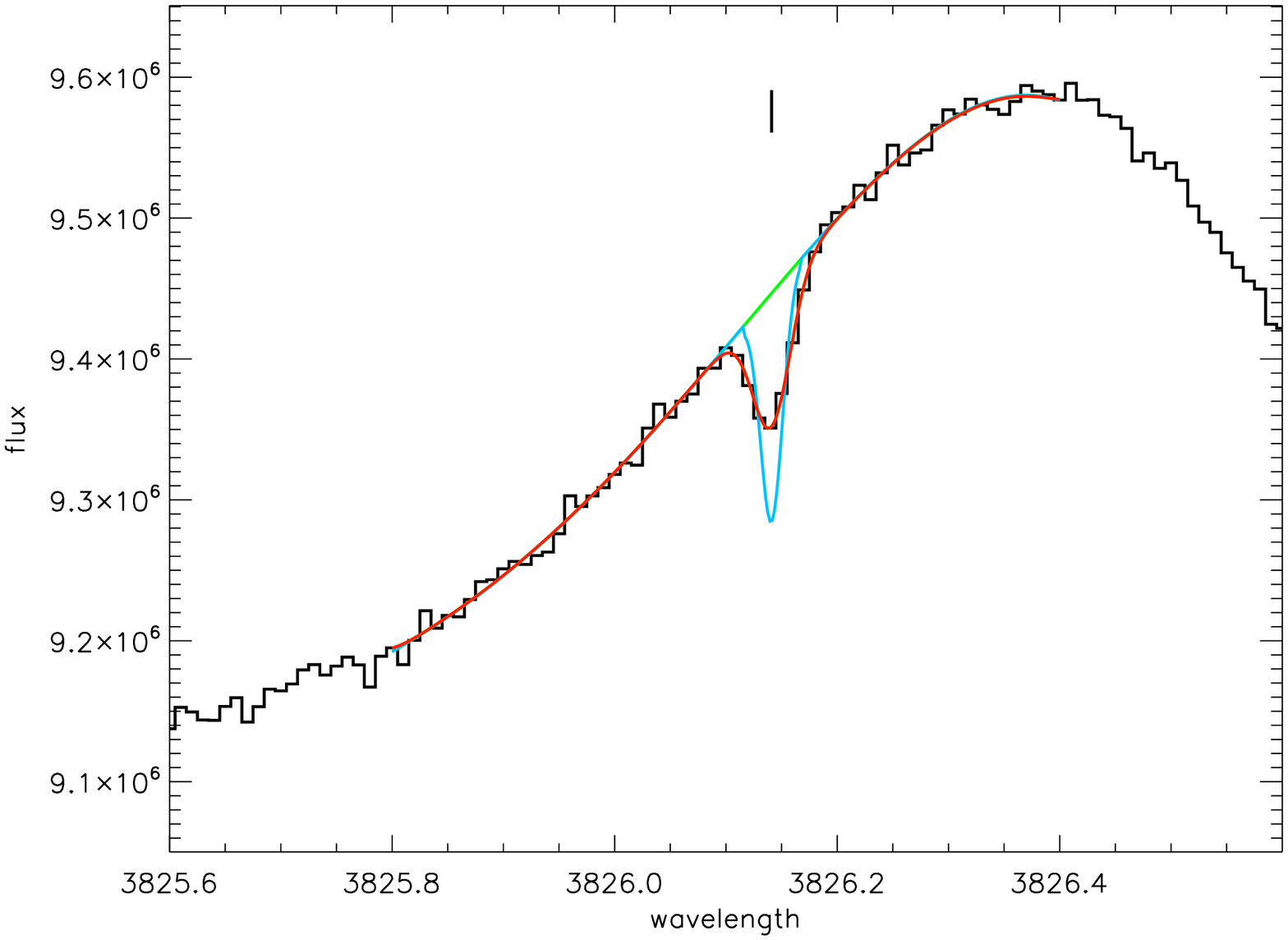}    
\includegraphics[trim=0.cm 0.cm 0.cm 0.75cm,angle=90,clip=true,width=0.33\textwidth]{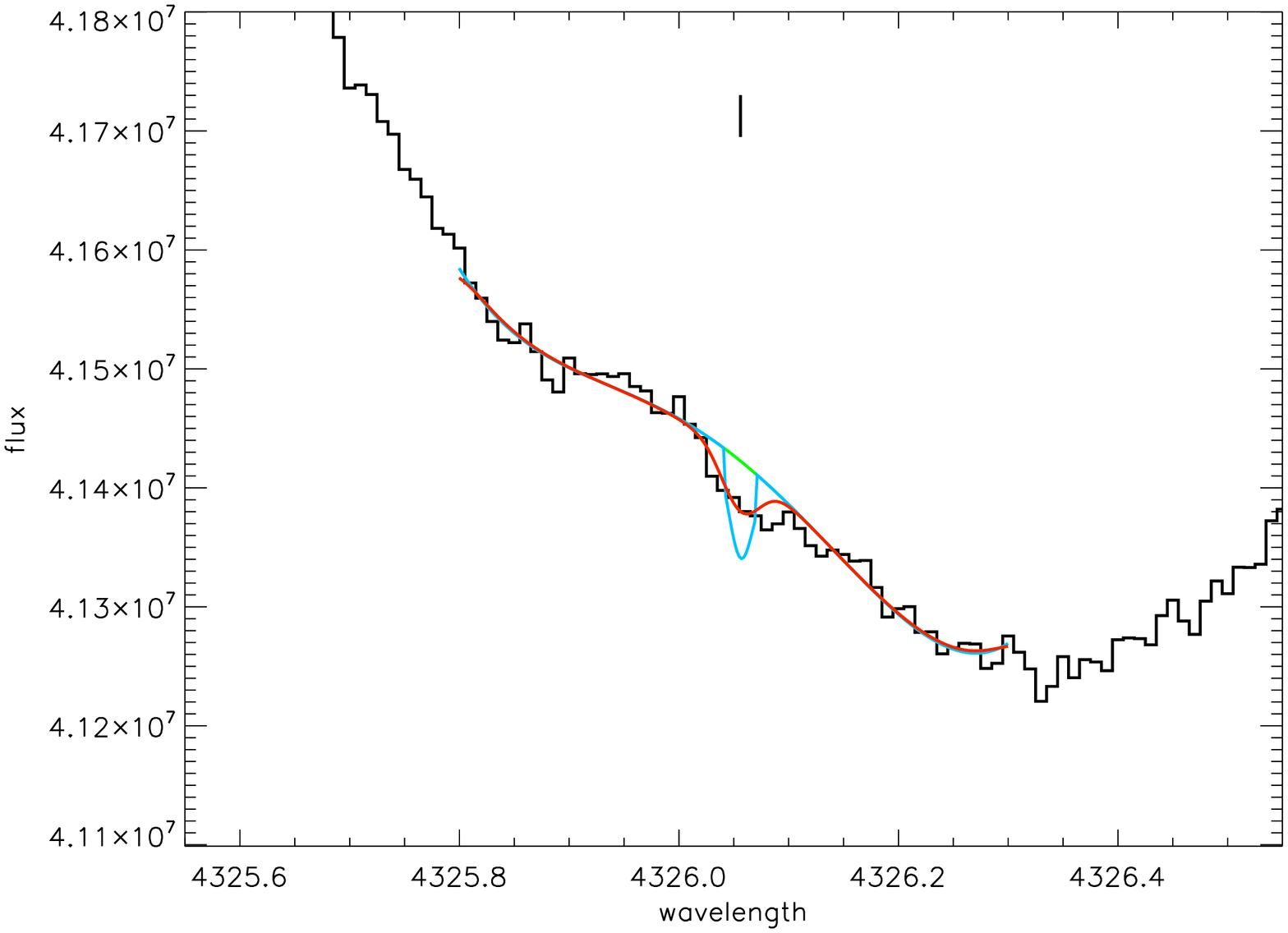}   
\hspace{0.72\textwidth}\\
\end{minipage}
\caption[]{
Simultaneous fits of the strongest Fe\,I lines from excited levels.
{\bf First column:} The 416 cm$^{-1}$, 704 cm$^{-1}$ and 888 cm$^{-1}$ excited levels (from top to bottom). 
{\bf Second column:} The 978 cm$^{-1}$, 6928 cm$^{-1}$ and 7377 cm$^{-1}$ excited levels (from top to bottom).
{\bf Third column:} The 11976 cm$^{-1}$, 12561 cm$^{-1}$ and 12969 cm$^{-1}$ excited levels (from top to bottom).
The stellar continuum is plotted with a green line, the intrinsic absorption profile with a blue line, and the 
absorption profile convolved with the HARPS instrument profile with a red line. The gas responsible for the 
absorption lines is defined as a unique component with the same radial velocity and line width ($b$ value) 
for all the excited levels; the column densities of each excited level are independent.
The positions of the absorption lines are indicated by a tick at +20.41 km/s radial velocity.
}
\label{Strongest_FeI_excited_fits}
\end{figure*}

\subsection{Circumstellar gas in the Fe\,I excited lines}
\label{FeI_excited_lines}

\subsubsection{Fe\,I* and Fe\,I** lines}

Following the same process as in Sect.~\ref{FeI_lines}, 
we analyzed the three strong lines from the two first excited levels at 416 cm$^{-1}$ 
(noted Fe\,I* at 3856.371 and 3886.282~\AA ) and 704 cm$^{-1}$ (noted Fe\,I** at 3878.573~\AA ). 
The results are presented in Table~\ref{FeIs_FeIss_parameters}. 

We found that each of these three lines is at the same radial velocity of 20.50\,km/s. 
This velocity is significantly higher than the radial velocity measured for the Fe\,I line 
from the ground base level, V$_{\rm FeI}=20.19$\,km/s. 
It also appears that the widths of the line for excited levels seem to be larger 
than those of the lines from ground base level, but the difference is barely significant.

\begin{table}
\centering
\caption{Same as Table~\ref{FeI_parameters} for the two Fe\,I* and one Fe\,I** lines}
\begin{tabular}{cccc}
 \hline
  V$_{\rm FeI*,FeI**}$   & b$_{\rm FeI*,FeI**}$ & N$_{\rm FeI*}$  & N$_{\rm FeI**}$ \\
  (km/s) & (km/s)   &   ($\times$10$^{12}$ cm$^{-2}$) &   ($\times$10$^{12}$ cm$^{-2}$) \\
 \hline
  20.50$\pm 0.10$  & 1.25$\pm 0.38$ & 0.46$\pm 0.04$  & 0.25$\pm 0.05$     \\
 \hline
\end{tabular}
\label{FeIs_FeIss_parameters}
\end{table}

\subsubsection{All the FeI lines from excited levels}

Here we assumed that all the detected absorptions in FeI lines from excited levels arise from a single medium. 
We thus used only a single absorption component to fit the profiles of all Fe\,I lines from the excited levels 
listed in Table~\ref{observations}. 
The fits of the strongest lines of each excited level are shown in the Fig.~\ref{Strongest_FeI_excited_fits}, 
excluding the lines from the 7728~cm$^{-1}$, 7986~cm$^{-1}$ ,
and 8155~cm$^{-1}$ excited levels 
because they are too weak to be visible in such a plot. 

Using a simultaneous fit to all the identified Fe\,I lines from excited levels, we found \\ 
\\
\centerline{V$_{\rm Exc}$~$=$~20.41~$^{+0.03}_{-0.05}$~km/s,}\\
\centerline{b$_{\rm Exc}$~$=$~1.01~$\pm$~0.06~km/s.}\\

The width of the lines is found to be narrow with $b\sim 1$~km/s. 
This strengthens the hypothesis that the entire absorption from the excited levels 
arises from a single component within the same circumstellar medium at the same radial velocity.
When we acknowledge that a fraction of the line widths is due to thermal broadening, 
the measured $b$ value provides an estimate for the maximum temperature of the gas, which is given by 
$T^{\rm Max}_{\rm Exc}$=$b_{\rm Exc}^2*m_{Fe}/(2k)$. 
We find  $T^{\rm Max}_{\rm Exc}$~$<$~3700~$\pm$~400~K.

\begin{table}
\centering
\caption{Column densities of the 12 excited levels of the FeI lines. $\sigma$ represents the significance of the detection.}
\begin{tabular}{rccc}
 \hline
 Excited   & N   & Error (+/-) & $\sigma$ \\
  level   & (cm$^{-2}$) & (cm$^{-2}$) & \\
 (cm$^{-1}$)   &  &  & \\
 \hline
416 & 0.393$\times$10$^{12}$ & (+0.011/-0.011)$\times$10$^{12}$ & 36. \\
704 & 0.220$\times$10$^{12}$ & (+0.010/-0.010)$\times$10$^{12}$ & 22. \\
888 & 0.857$\times$10$^{11}$ & (+0.081/-0.079)$\times$10$^{11}$ & 11. \\
978 & 0.296$\times$10$^{11}$ & (+0.099/-0.080)$\times$10$^{11}$ & 3.0 \\
6928 & 0.819$\times$10$^{11}$ & (+0.021/-0.019)$\times$10$^{11}$ & 41. \\
7377 & 0.303$\times$10$^{11}$ & (+0.016/-0.015)$\times$10$^{11}$ & 20. \\
7728 & 0.720$\times$10$^{10}$ & (+0.400/-0.390)$\times$10$^{10}$ & 1.8 \\
7986 & 0.106$\times$10$^{11}$ & (+0.025/-0.026)$\times$10$^{11}$ & 4.2 \\
8155 & 0.451$\times$10$^{10}$ & (+0.270/-0.270)$\times$10$^{10}$ & 1.7 \\
11976 & 0.469$\times$10$^{10}$ & (+0.049/-0.052)$\times$10$^{10}$ & 9.2 \\
12561 & 0.205$\times$10$^{10}$ & (+0.055/-0.011)$\times$10$^{10}$ & 19. \\
12969 & 0.129$\times$10$^{10}$ & (+0.019/-0.023)$\times$10$^{10}$ & 5.6 \\
\hline
\end{tabular}
\label{FeI_levels_N}
\end{table}

The measured column densities of the 12 Fe\,I excited levels are given in Table~\ref{FeI_levels_N}. 
For ten of the listed excited levels, the signatures are detected at more than 3$\sigma$. 
For the levels from 7728~cm$^{-1}$ and from 8155~cm$^{-1}$, the detections are at a significance lower than 2$\sigma,$
and the corresponding column densities can be considered as upper limits. 
The measured column densities cover more than two orders of magnitude.

\subsection{Two-component Fe\,I ground base level}
\label{FeI_2comp_section}

The measured radial velocity of the gas component producing the absorption in all these excited Fe\,I lines, 
V$_{\rm Exc}=20.41~^{+0.03}_{-0.05}$~km/s, is different at 4$\sigma$ from the radial velocity 
measured in the two Fe\,I lines from the ground base level, V$_{\rm FeI}=20.19\pm 0.05$ km/s. 
Nonetheless, the gas observed in the excited levels must 
also include a population of Fe\,I atoms at the ground base level. 
Therefore the measured difference in radial velocity supports the idea that the absorptions 
in the Fe\,I lines from the ground base level can have two components: one at the same radial velocity as
the absorption from the excited levels, and a second component at a lower radial velocity.

\begin{figure}
\begin{minipage}[b]{\columnwidth}         
\includegraphics[trim=0.cm 0.cm 0.cm 0.75cm,angle=90,clip=true,width=0.99\columnwidth]{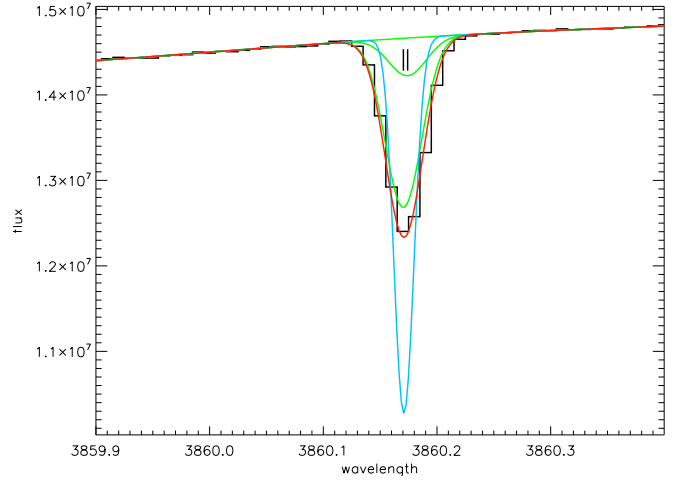}    
\hspace{0.18\columnwidth}\\
\includegraphics[trim=0.cm 0.cm 0.cm 0.75cm,angle=90,clip=true,width=0.99\columnwidth]{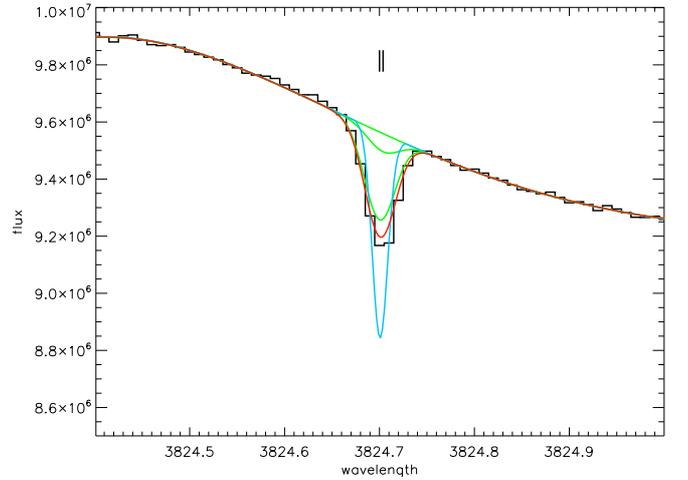}  
\hspace{0.18\columnwidth}\\
\end{minipage}
\caption[]{Same as Fig.~\ref{FeI_HARPS} with two-component fits to the line absorption profiles. 
The positions of the two components are shown by two small vertical tick marks. 
The absorption profiles of each of the two components are shown by green lines. 
The redward component (1) is at the same radial velocity as the absorption from all excited levels. 
The blueward component (2) has a higher column density, but is detected only in the FeI lines from the ground base level. 
}
\label{FeI_2comp}
\end{figure}

We therefore fit the profile of the Fe\,I lines from ground base level with two components: 
a redward component (1) using the same radial velocity as the velocity found for the absorptions from all excited levels, 
and a blueward component (2) with a radial velocity let free to vary in the fitting procedure (Fig.~\ref{FeI_2comp})
The results are the following: \\
\\
\centerline{$V^{1}_{\rm FeI}$~=~20.41~km/s (fixed),} \\
\centerline{$b^{1}_{\rm FeI}$~=~1.01~km/s (fixed),} \\
\centerline{$N^{1}_{\rm FeI}$~=~(0.956~$^{+0.002}_{-0.059}$)$\times$10$^{12}$~cm$^{-2}$,}\\
\\
\centerline{$V^{2}_{\rm FeI}$~=~20.07~$^{+0.02}_{-0.01}$~km/s,}\\
\centerline{$b^{2}_{\rm FeI}$~=~0.59~$^{+0.04}_{-0.01}$~km/s,}\\
\centerline{$N^{2}_{\rm FeI}$~=~(1.57~$^{+0.06}_{-0.01}$)$\times$10$^{12}$~cm$^{-2}$,}\\
\\
\\
\centerline{$N^{\rm Total}_{\rm FeI}$~=~(2.53~$^{+0.06}_{-0.07}$)$\times$10$^{12}$~cm$^{-2}$,}\\
\centerline{$\Delta V$~=~0.34~$^{+0.04}_{-0.05}$~km/s.}\\

The velocities of the two components V$^{1}_{\rm FeI}$ and V$^{2}_{\rm FeI}$ are found to be significantly different. 
The highest column density in Fe\,I is found to be in the second component (2).

The presence of two components is supported by the goodness-of-fit improvement 
when using the two-components model. 
With the addition of a second component, the $\chi^2$ decreases from 120.3 (with 139~degrees of freedom) 
to 114.9 (with 138~degrees of freedom).
The Akaike information criterion (AIC) decreases by 3.4 with the second model, 
which shows that the presence of a second component is definitely favored (the BIC is also lower). 

When $V^1$ and $b^1$ were left free to vary in the fit, 
the $\chi2$ decreases only to 112 with similar values for all parameters, despite the two new degrees of freedom. 
This confirms that the detected second component in the Fe\,I ground base level 
is likely the expected counterpart of the absorption seen in the excited levels,  
with the same physical properties ($V$ and $b$) for both excited and ground base levels. 

\section{Discussion}
\label{Discussion}
 
\subsection{Fe\,I excited levels} 
\label{The FeI excited levels} 

The Fe\,I excited levels can be populated either by radiation 
or by collisions if the medium is dense enough. Therefore, the Fe\,I excited levels signatures 
are not observed in the interstellar medium (ISM) absorption lines because the volume density of the ISM in the vicinity of the Sun 
 and the exciting photon flux are too low. 
The Fe\,I ground base level lines are also not usually observed in the ISM for another reason: 
iron atoms are easily ionized in the ISM UV flux. As a result, all the Fe\,I lines detected here 
arise from the circumstellar medium of \bp .

\begin{figure}[tb]
\begin{minipage}[b]{\columnwidth}         
\includegraphics[trim=0.cm 0.cm 0.cm 0.75cm,angle=0,clip=true,width=0.95\columnwidth]{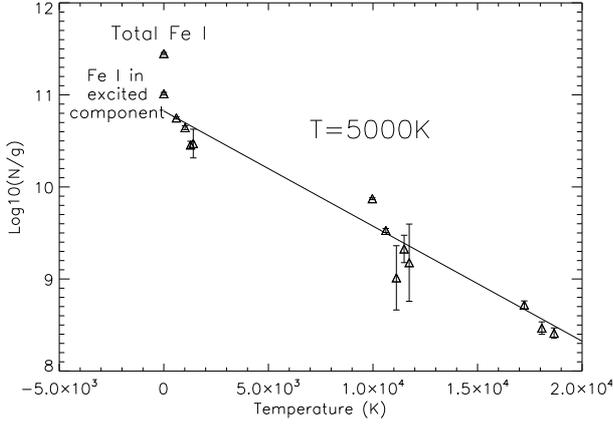}   
\hspace{0.18\textwidth}\\
\end{minipage}
\caption[]{Excitation diagram of the level populations with log$_{10}$($N$/$g$) as a function of the level energy 
in Kelvin with 1$\sigma$ error bars.
The least-squares fit is shown as a straight solid line and corresponds to a 5000\,K thermodynamic equilibrium distribution. 
The disagreement between this line and the observations shows that the observations do not correspond to LTE conditions.
}
\label{log_Nsg_T}
\end{figure}

\begin{table}
\centering
\caption{Sensitivity to the stellar flux of the different excited levels, 
as shown by the sum of the oscillator strengths of their lines, 
below or above the Balmer discontinuity (at about 3700\AA ), {\it b}BD or {\it a}BD, respectively. 
{\it N}$_{\rm {\it b}BD}$ and {\it N}$_{\rm {\it a}BD}$ are the numbers of lines for each category, and 
$\Sigma _{\rm {\it b}BD}{\rm f}$ and $\Sigma _{\rm {\it a}BD}{\rm f}$ are the sums of the oscillator strengths. 
}
\begin{tabular}{rrcrcc}
 \hline
  E$_{\rm i}$~~~ &  {\it N}$_{\rm {\it b}BD}$ & $\Sigma _{\rm {\it b}BD}{\rm f}$ &  {\it N}$_{\rm {\it a}BD}$ & 
  $\Sigma _{\rm {\it a}BD}{\rm f}$  & ~~$\Sigma _{\rm {\it b}BD}{\rm f}~/$ \\
  (cm$^{-1}$) &  &  & & & $\Sigma _{\rm {\it a}BD}{\rm f}$\\ 
 \hline
 0 & 22 & 1.3964 & 3 & 0.06763 & 20.65 \\     
  416 & 29 & 1.3245 & 5 & 0.06730 & 19.68 \\      
  704 & 33 & 1.3860 & 5 & 0.06491 & 21.35 \\      
  888 & 29 & 1.4609 & 4 & 0.06273 & 23.29 \\      
  978 & 14 & 1.5324 & 2 & 0.06410 & 23.91 \\      
\hline
  6928 & 27 & 0.63980 & 3 & 0.31334 & 2.042 \\      
  7377 & 34 & 0.65591 & 6 & 0.30843 & 2.127 \\      
  7728 & 36 & 0.68351 & 7 & 0.29204 & 2.340 \\      
  7986 & 32 & 0.65986 & 8 & 0.27952 & 2.361 \\      
  8155 & 25 & 0.68376 & 6 & 0.26621 & 2.568 \\      
\hline
  11976 & 9 & 0.04452 & 8 & 0.73459 & 0.0606 \\      
  12561 & 13 & 0.04715 & 10 & 0.72093 & 0.0654 \\      
  12969 & 10 & 0.04980 & 8 & 0.73392 & 0.0679 \\      
\hline
\end{tabular}
\label{Low_flux_FeI}
\end{table}

In an excitation diagram [$E_K$,log$_{10}$($N$/$g$)] where for each considered excited level, $E_K$ is the level energy expressed in Kelvin, $N$ the column density and $g$ the g-factor ($g$=2$J$+1, see Table~\ref{levels}), the population of the excited levels at local thermodynamic equilibrium (LTE) follow a linear relationship whose slope is linked to the temperature of the medium. 
This diagram is shown in Fig.~\ref{log_Nsg_T} for the Fe\,I excited level column density evaluations. Their distribution is not linear, showing that LTE conditions are not met in the considered medium where the Fe\,I lines are observed. The least-squares linear fit is shown. It corresponds to a temperature of about 5000\,K, quite different from the effective temperature of the \bp\ photosphere (T$_{eff}\sim~$8000\,K). 
This shows that the populations of the different levels are not only controlled by radiation, but also by collisions. This argues for an equilibrium that is neither radiative nor collisional, but somewhere in between radiative and collisional regimes,  as described for instance by{\it } Viotti (1976; see Sect.~\ref{Possible regimes}). 

\begin{figure}
\includegraphics[width=\columnwidth]{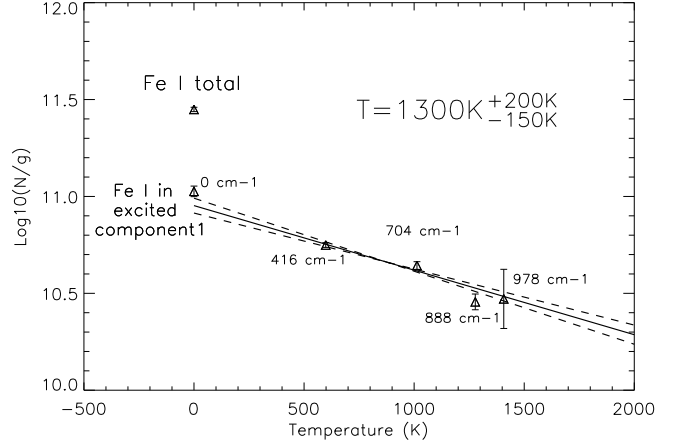} 
\includegraphics[width=\columnwidth]{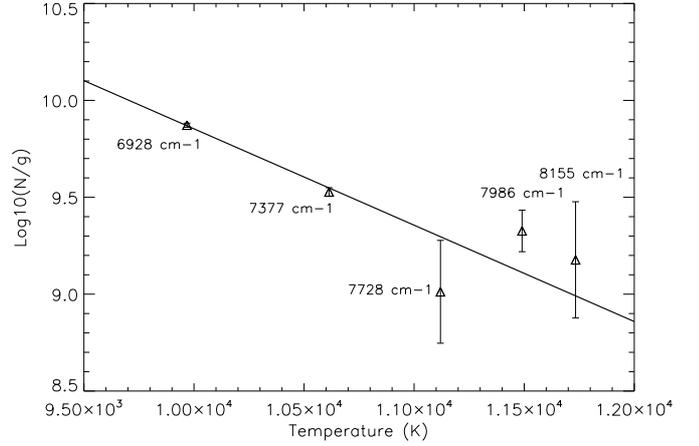} 
\includegraphics[width=\columnwidth]{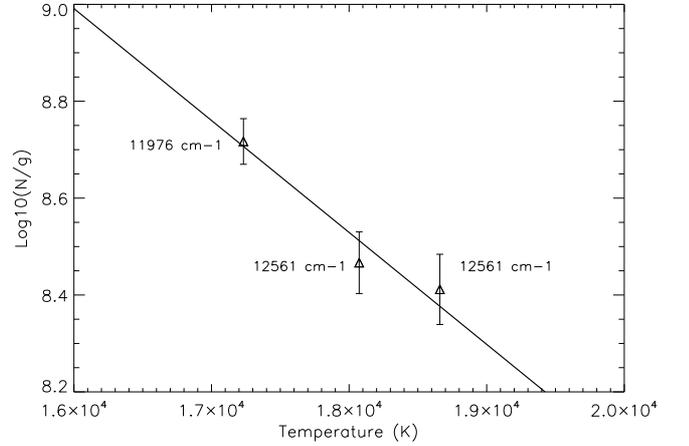}        
\caption[]{Same as Fig.\ref{log_Nsg_T}, but in three separate plots corresponding to the three domains. 
{\bf Top:} Low-energy levels. {\bf Middle:} Medium-energy levels. {\bf Bottom:} High-energy levels.
Least-squares fits 
of all available levels  
are shown as solid straight lines. For the low{\bf } energy levels, the 0\,cm$^{-1}$ level corresponding to component 1 is not included in the fit (see text).
Limits on the least-squares slope
are shown as dashed lines for the low-energy levels, centered over the weighted average of the four excited levels. 
The population of the low-energy levels follow a Boltzmann distribution, which is consistent with an excitation 
temperature of about 1300\,K.
}
\label{Detailed_FeI_diagrams}
\end{figure}

In the energy diagram (Fig.~\ref{log_Nsg_T}), the energy levels are clustered within three groups. 
We recall that the slope corresponding to 8000~K equilibrium conditions should be smaller than the slope of the 5000~K conditions, therefore the lower energy level column densities are relatively too high, while the higher energy levels are relatively too low.

Because the \bp\ spectrum is not a perfect 8000~K black body, this behavior can be explained qualitatively. The three groups of energy levels are supplied by atomic transitions that are
distributed along the stellar spectrum. Strong flux reductions, such as the deep Balmer lines and the Balmer discontinuity (BD) of an A5 spectrum below 3700 \AA, will effectively underpopulate certain energy levels.
For each energy level, we calculated the sum of all the oscillator strengths 
(with f~$>$10$^{-3}$) of all lines between 2000\AA\ and 9000\AA\ (where most of the Fe\,I lines occur). 
The number of lines counted {\it \textup{below}} ($N_{\rm {\it b}BD}$) and {\it \textup{above}} ($N_{\rm {\it a}BD}$) 
the Balmer discontinuity, as well as the sum of their oscillators strengths, 
$\Sigma _{\rm {\it b}BD}{\rm f}$ and $\Sigma _{\rm {\it a}BD}{\rm f,}$ are given in Table~\ref{Low_flux_FeI}. 
The ratio $\Sigma _{\rm {\it b}BD}{\rm f}$~/~$\Sigma _{\rm {\it a}BD}{\rm f}$, 
which represents the sensitivity of the considered level to the stellar flux, is also listed. 
These ratios are significantly different in the three groups of energy levels: 
$\sim$20 for $E_{\rm i}$~$<$~1000~cm$^{-1}$, $\sim$~2 for 6000~cm$^{-1}$~$<$~$E_{\rm i}$~$<$~9000~cm$^{-1}$ 
, and $\sim$~0.06 for $E_{\rm i}$~$>$~11000~cm$^{-1}$. 
In these three domains, the radiative contribution is expected
to be very different, as described below.\\ 
1) Group of High-energy levels above 11000~cm$^{-1}$. 
The highest energy levels have the highest photon-pumping efficiency 
because their spectroscopic lines lie higher than the Balmer discontinuity. 
For these levels, the radiative contribution is expected to be the dominant mechanism.\\
2) Group of Medium-energy levels, from 6000~cm$^{-1}$ to 9000~cm$^{-1}$. 
For the levels in the group at intermediate energy, the photon-pumping efficiency has a medium value 
because the strongest lines of these levels are still present below the Balmer discontinuity. 
Here the radiative contribution should be less than for the higher energy levels. \\
3) Group of Low-energy levels, below 1000~cm$^{-1}$. 
The efficiency of the photon pumping in populating the low-energy levels is the lowest 
because their spectroscopic lines are mostly below the Balmer discontinuity. 
There the population of the levels by radiation is relatively less important 
and the dominant process could thus possibly be collisions.\\

The relative population of different energy levels can be analyzed within each group of energy levels: 
the group below 2000\,K ($<$1000~cm$^{-1}$), the group around 11000\,K (from 6000~cm$^{-1}$ to 9000~cm$^{-1}$), 
and the group around 18000\,K ($>$11000~cm$^{-1}$).
The energy diagrams for each group are given in Fig.~\ref{Detailed_FeI_diagrams}. 
It is noteworthy that the column density measured in component (1) of the ground base level is consistent 
with the column density extrapolated from the Boltzmann distribution in the energy diagram when we consider 
only the excited levels in the low-energy group (upper panel of Fig.~\ref{Detailed_FeI_diagrams}), 
although they are one order of magnitude lower than the total Fe\,I column density. 
This strengthens the idea that component (1) in the ground base level is the counterpart of the gas 
detected in the exited levels. 

From the alignment of the level populations, we can derive a characteristic temperature   
for the low-energy levels of $1300^{+200}_{-150}$\,K.
Because the population in the low-energy levels is likely dominated by collisions, 
we can hypothesize that this temperature is a proxy for the measurement of the iron gas temperature.
The b value of the same medium is evaluated to be $b_{\rm Exc}$=$1.01$$\pm$0.06\,km/s, which corresponds 
to a maximum temperature of $T_{\rm Exc}$$<$3700$\pm$400\,K.
Assuming $T_{\rm Exc}$=1300\,K, we can derive the corresponding turbulent broadening $\xi_{\rm Exc}$ to be\\
\\
\centerline{ 
$\xi_{\rm Exc}$~=~$0.80~\pm 0.07$~km/s.
}

This new estimate of the turbulent broadening in the stable circumstellar gas yields a new constraint on any line width, which must be larger than $\sim$~0.8~km/s, for any species. Of course, this constraint applies only where collisional equilibrium is reached and does not applie, for instance, in component (2) of Fe\,I described above, for which we found b~$\sim$~0.6\,km/s (see Section~\ref{FeI_2comp_section}). In this last case, the broadening is likely dominated by radiation pressure on the Fe\,I atoms, as suggested by Brandeker (2011).

Most importantly, the temperature $\sim$1300~K is found to be in striking agreement with the iron condensation temperature,
which is of about 1350\,K (Lodders 2003). This temperature could thus be considered as the sublimation temperature of grains, possibly drifting toward the star under the Poynting-Robertson effect, gas drag, or any other mechanism. 
Such grains can be produced by the evaporation of the exocomets that are regularly observed in the \bp\ system (Kiefer et al. 2014), 
and can be the source of the Fe\,I seen in the stable circumstellar gas. 
In this scenario, the location of  the observed Fe\,I atoms should be at the sublimation radius around the star, that is, 
at a distance of about 38\,R$_{\rm Star}$ or $\sim$0.2\,au from the star, where the equilibrium temperature of grains is expected 
be about 1350~K.

\begin{figure}[tb]
\begin{minipage}[b]{\columnwidth}         
\includegraphics[trim=0.cm 0.cm 0.cm 0.75cm,angle=0,clip=true,width=0.95\columnwidth]{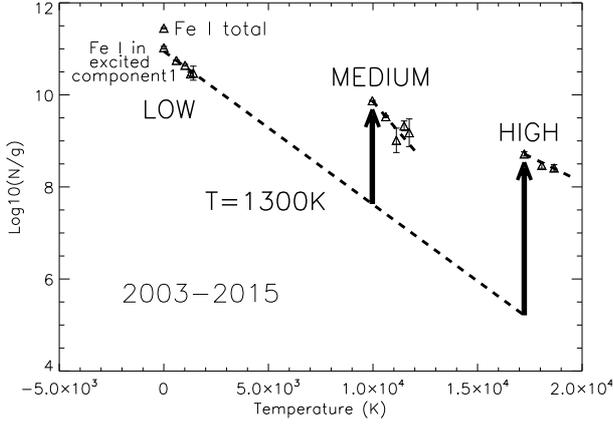}       
\hspace{0.18\textwidth}\\
\end{minipage}
\caption[]{Same as Fig.\ref{log_Nsg_T}.
The 1300\,K thermodynamic equilibrium distribution corresponding to the lower energy levels is shown by a straight dashed line. 
The disagreement between this line and the observations related to the higher energy levels shows that indeed LTE is not reached for all levels. Table~\ref{Low_flux_FeI} explains at least qualitatively why such a situation is encountered, the two higher groups of energy levels are increasingly sensitive to radiation from the stellar photospheric flux, as shown by the two vertical jumps underlined by the two vertical arrows.
Conversely, the lowest levels have five well-aligned levels and must be
less sensitive to radiation and thus are mostly populated by another mechanism, likely collisions (see text).
}
\label{log_Nsg_T_conclusion}
\end{figure}

\subsection{Physical conditions within the considered medium} 
\label{Physical conditions} 

In Fig.~\ref{log_Nsg_T_conclusion} we show all the Fe\,I excited levels, along with the 1300~K linear fit 
to the low-energy levels. 
If LTE conditions were met, the population of the medium-{\bf } and high-energy levels 
would be distributed along that same line. This is obviously not the case, 
and as is shown by the vertical arrows, a jump is needed to reach the observed population. 
In agreement with Table~\ref{Low_flux_FeI}, the radiation contribution to the equilibrium increases 
for groups on higher excitation levels. The higher the group, the larger the jump. 
As the contribution of radiation increases, the level distribution departs from a local linear distribution, 
showing evidence of non-LTE conditions for these groups.

However, we can still question the role of radiation in the low-energy level distribution.

\subsubsection{Electronic volume density} 
\label{Electronic volume density} 

One of the main effects of radiation is photoionization and thus the liberation of electrons in the surrounding medium. 
Measuring the electronic density will allow us to constrain the regime (collisional or radiative) of the medium at 38\,$R_\star$
from \bp\ in Sect.~\ref{Possible regimes}. 
Lagrange et~al. (1995) showed that the observed ratio N(FeI)/N(FeII)$<$$10$$^{-2}$, balanced by competing effects of radiation, 
collisions, and recombinations,
requires the following condition at a distance r from the star: 4.5$\times$\!\,$10^3/r^2$\!\,$>$\!\,$N_e$\!\,$\times$\!\,$T_e^{-0.89}$, 
where $N_e$ and $T_e$ are the electronic density and temperature of the medium, and $r$ is in AU. 
Applying this relation to a medium at 1300~K at 0.2~AU from the star, we find that $N_e$\!\,$<$\!\,$6.7$\!\,$\times$\!\,$10^7$~cm$^{-3}$. 
However, the Fe\,I lines are within a narrow spectral domain at about 20.4\,km/s, while the Fe\,II absorption 
is observed at about 22\,km/s (Lagrange et al.\ 1995); the N(FeI)/N(FeII) ratio should hence 
be estimated using the column density of FeII at 20.4\,km/s (or 20.9\,km/s when we take into account 
the 0.5~km/s shift between FeII and FeI; Brandeker et al.\ 2011).
Lagrange et al. (1995) also evaluated the FWHM of Mn\,II and Zn\,II, which are 5.6 and 5.5\,km/s, respectively. 
Thus only a small overlap exists between the 20.4\,km/s spectral position and the blue edge of the Fe\,II line. 
This leads to a rough estimate of the FeII column density in the same component as in FeI of about 
10$^{-2}$ times the total column density measured by Lagrange et al.\ (1995). 
The column density ratio is thus more likely N(FeI)/N(FeII)$\la$1, 
which yields an upper limit for the electronic density of $N_e$\!\,$\la$\!\,$7$\!\,$\times$\!\,$10^5$~cm$^{-3}$. 
Combined with the condition given by Kondo \&\ Bruhweiler (1985) that $N_e$\!\,$\ge$\!\,$10^3$~cm$^{-3}$ 
to explain the presence of excited levels of C\,I and Fe\,I, we find that the electron density should be~\smallskip\\
\centerline{$10^3$~cm$^{-3} \le N_e \la 7\times10^5$~cm$^{-3}$}
\\

\subsubsection{Possible equilibrium regimes} 
\label{Possible regimes} 

Following Viotti (1976), three possible regimes could be observed in a stellar vicinity: collisional, radiative and nebular.

Viotti (1976) showed in particular that these three regimes could be visualized in a $[W,N_eT_e^{1/2}]$ diagram, 
where $W$ is the dilution factor that measures the density of the stellar radiation in the stellar environment, 
and $N_eT_e^{1/2}$ represents the rate of electron impacts.

With $W$ = $1/2\{1-[1- ($R$_{\rm Star}/$R$)^2]^{1/2}\},$ we can evaluate that the dilution factor at R~=~$38\times$R$_{\rm Star}$ is log$W$~=~-3.76. With the previous constraints found on $Ne$ and assuming $T_e$~=~1300~K, we should have~
$1.44\le~$log$N_eT_e^{1/2}~<4.27$.

The $[W,N_eT_e^{1/2}]$ diagram (Viotti 1976~; Fig~2) shows that the evaluated 
conditions locate the considered \bp\ circumstellar medium in the radiative regime. 
However, Viotti (1976) computed his diagram for Be stars, which
are more energetic in the UV than \bp , and indicated 
in the diagram the positions of one more UV-energetic nova as well as the conditions in the photosphere 
of less UV energetic B-type stars (dwarf, B\,V, and supergiants, B\,I). 
We conclude that in the environment of less UV-excited stars like \bp\ (an A5V star), 
the position is shifted toward the right part of the diagram. 
This would place the medium of the observed excited Fe\,I lines in the radiative to collisional transition region, which is consistent with our evaluation from the excitation diagrams 
(see Fig.~\ref{log_Nsg_T_conclusion}).

\section{Conclusion}

The present study of the large set of spectra of \bp\ obtained
with the HARPS instrument allowed a detailed analysis 
of the characteristics of the Fe\,I component and deriving a possible scenario for the origin of Fe\,I atoms 
observed in the ``stable'' circumstellar gas disk.\\
\\
The main conclusions are~listed below.
\begin{itemize}
\item Fe\,I is detected from the ground base level up to the 12969~cm$^{-1}$ energy level.
\item Fe\,I from the ground state is seen to be present within two components, one at 20.07~km/s 
and the other at 20.41~km/s radial velocity. The component at the lowest radial velocity 
is blueshifted relative to the main circumstellar gas; this blueshift gas can be 
dragged away by radiation pressure, as suggested by Brandeker (2011).
\item The component at 20.41~km/s is common to all Fe\,I excited states. 
This component can be directly linked to source of Fe\,I atoms.
\item The distributions of the low-energy levels are aligned in the excitation diagram and 
correspond to a temperature of about $\sim$~1300~K for the gas in the 20.41\,km/s component.
\item This temperature is similar to the Fe\,I sublimation temperature. 
We therefore conclude that the source of iron atoms in the inner circumstellar disk may be 
the evaporation of dust grains at about 0.2\,AU or 38\,R$_{\rm Star}$ from the stellar surface.
\end{itemize}

Although a complete non-LTE study needs to be made over the whole set of the excited Fe\,I level 
populations to completely identify the physical conditions in the medium, 
following Viotti (1976), we showed that the medium is likely in a transition regime 
between a radiative and a collisional regime. 
Further modeling is needed to confirm these preliminary evaluations.

\begin{acknowledgements}
We warmly thank F.~Bouchy and the HARPS team for fruitful discussions on the subject of the present article. 
Support for this work was provided by the CNES. 
We acknowledge the support of the French Agence Nationale
de la Recherche (ANR), under program ANR-12-BS05-0012 "Exo-Atmos".
VB and DE acknowledge the financial support of the SNSF.
\end{acknowledgements}

\end{document}